%% file: selfenergy_prd.tex
\documentclass[
superscriptaddress,
nofootinbib,
 amsmath,amssymb,
 aps,
 prd,
floatfix,
twocolumn
]{revtex4}

\usepackage{graphicx}
\graphicspath{{./pic/carlo/}{./pic/fra/}{./pic/pp/}}
\usepackage{bm}  
\usepackage[utf8]{inputenc}
\input{config}

\newcommand{\ra}{\rangle}

\newcommand{\nn}{\nonumber}

\newcommand{\Wthree}[6]{\left(\begin{array}{ccc} #1 & #2 & #3 \\ #4 & #5 & #6 \end{array}\right)}
\newcommand{\Wfour}[9]{\left(\begin{array}{cccc} #1 & #2 & #3 & #4 \\ #5 & #6 & #7 & #8 \end{array}\right)^{(#9)}}
\newcommand{\Wsix}[6]{\left \{ \begin{array}{ccc} #1 & #2 & #3 \\ #4 & #5 & #6 \end{array}\right \} }

\newcommand{\slnext}{\texttt{sl2cfoam-next}}

\usepackage[colorinlistoftodos,prependcaption,textsize=footnotesize]{todonotes}

\begin{document}

\title{Numerical analysis of the self-energy in covariant LQG}

\author{Pietropaolo Frisoni}
\email{pfrisoni@uwo.ca}
\affiliation{Dept.\,of Physics \& Astronomy, Western University, London, ON N6A\,3K7, Canada}

\author{Francesco Gozzini}
\email{gozzini@cpt.univ-mrs.fr}
\affiliation{Aix Marseille University, Universit\'e de Toulon, CNRS, CPT, 13288 Marseille, France}

\author{Francesca Vidotto}
\email{fvidotto@uwo.ca}
\affiliation{Dept.\,of Physics \& Astronomy, Western University, London, ON N6A\,3K7, Canada}
\affiliation{Dept.\,of Philosophy and Rotman Institute, Western University, London, ON N6A\,3K7, Canada}

\begin{abstract}
\noindent We study numerically the first order radiative corrections to the self-energy, in covariant loop quantum gravity. We employ the recently developed \slnext{} spinfoam amplitudes library, and some original numerical methods.  We analyze the scaling of the divergence with the infrared cutoff, for which previous analytical estimates provided widely different lower and upper bounds. Our findings suggest that the divergence is approximately linear in the cutoff. We also investigate the role of the Barbero-Immirzi parameter in the asymptotic behavior, the dependence of the scaling on some boundary data and the expectation values of boundary operators.
\end{abstract}

\maketitle

{
  \hypersetup{linkcolor=black}
}


\section{Introduction}
\label{sec:introduction}

The spinfoam, or covariant, formulation of loop quantum gravity \cite{Perez:2012wv
}, is based on the definition of truncated amplitudes that define the background independent dynamics of quantum spacetime \cite{Oeckl:2003vu,Rovelli2004} in a way similar to the Feynman graphs' definition of the dynamics of quantum field theory. In a Feynman graph, radiative corrections appear as loops (set of propagators forming a circle) and may give rise to ultraviolet divergences.  In a spinfoam, radiative corrections appear as bubbles (set of faces forming a sphere) and may give rise to infrared divergences when the cosmological constant is taken to be zero\footnote{The general definition of the spinfoam amplitudes includes a non-zero cosmological constant \cite{Fairbairn:2011aa,Han:2011nx,Haggard:2015yda} that plays the role of an infrared cutoff and makes the amplitudes finite.}. Here we investigate the first order correction to a single edge of the spinfoam, which is analog to the first order contribution to the self-energy in quantum field theory, hence are referred to as self-energy radiative corrections.  We use the version of the amplitude developed in \cite{Pereira:2007nh, Livine:2007vk, Livine:2007ya, Freidel:2007py, Engle2007,Engle:2007uq,Engle:2007wy, Kaminski:2009fm, Ding:2010fw} generally denoted as the EPRL amplitude. For reviews and notation, see \cite{ Rovelli:2014ssa,Rovelli2011c,Dona:2010cr}. 

The spinfoam infrared divergences have been studied analytically from a number of different perspectives 
\cite{dittrich_2016_continuum_limit, Dittrich_decorated_tensor_network, Bahr_Investigation_of_the_spinfoam,Bonzom_random_tensor_models, Benedetti_phase_transition_in_dually, Carrozza_renormalization_of, Geloun_functional_renormalization,Bonzom:2010zh, Bonzom:2011br, Baratin_melonic_Phase_Transition}. EPRL divergences have been studied in the Euclidean \cite{Perini:2008pd, Krajewski:2010yq} and Lorentzian theories \cite{Riello:2013bzw,Dona:2018infrared}. In particular the last two references by Riello and Don\'a provide lower and upper bounds to the degree of divergence of the self-energy, resulting in a wide window. In \cite{Riello:2013bzw}, Riello gives a detailed analysis of the critical point that contributes to the amplitude in the large-spin scaling limit, based on the techniques introduced in \cite{article_Barrett_etal_2010_lorentzian_spinfoam_amplitude}, in the case of non-degenerate geometric configurations. He finds a logarithmic divergence in the cutoff $K$. He then estimates in a final appendix that the contribution of the degenerate sector might be linear in $K$, hence dominating the total divergence, postponing a detailed analysis of this sector to future studies. On the other hand, in \cite{Dona:2018infrared}, Don\'a disregards the interference between the different terms in the sum, and this allows him to estimate an upper bound on the divergence that scales as $K^9$. He also proves that the upper bounds derived with his algorithm provide excellent estimates of different kinds of infrared divergences restricted to the 3D model, suggesting that this might be also true in 4D, underlining the need for a complete numerical analysis for future studies.

These two results leave a window of possibilities which spans more than nine powers of the cutoff. In this work, we address the issue numerically.  What we find is consistent with Riello's rough estimate: for small values of the Barbero-Immirzi parameter, the numerical analysis is compatible with a \emph{linear} divergence in $K$. 

The numerical calculation is possible thanks to the recently developed \slnext{} library \cite{Francesco_draft_new_code}. Our computations were performed on the clusters Cedar and Graham of Compute Canada (\href{https://www.computecanada.ca/}{www.computecanada.ca}) and on the Mesocentre of Aix-Marseille University (\href{https://mesocentre.univ-amu.fr}{mesocentre.univ-amu.fr}). The analysis and visualization of the computed data have been done with Julia \cite{bezanson2017julia} and Mathematica \cite{Mathematica}.  

In section \ref{sec:self_energy_diagram} we describe the structure of the self-energy amplitude. In section \ref{sec:derivation_BF_amp} we recall the calculation of the degree of divergence in topological $SU(2)$ BF model, which can be found exactly. This allows us to compare the numerical results with the analytical ones in the literature. In section \ref{sec:From_BF_to_EPRL} we define the EPRL amplitude. In section \ref{sec:Div_analysis} we describe the results of the numerical analysis 
of the EPRL 
self-energy amplitude. Finally, in the appendices we report some further numerical studies and consistency checks.

\newpage

\section{Self-energy} 
\label{sec:self_energy_diagram}
Consider two four-simplices joined by four tetrahedra. The boundary of this cellular complex is formed by two tetrahedra. In quantum gravity, this cellular complex can be associated to the transition from a single quantum of space to a single quantum of space, via a splitting into four virtual quanta, which then recombine \cite{Oeckl:2003vu,Rovelli2004}. 
Notice the similarity with the self-energy Feynman graph in a quantum field theory. This however is only an analogy: there is no notion of energy or radiation involved here. This graph should be understood as a higher order term in the spinfoam expansion.

\begin{figure}[h]
    \centering
    \includegraphics[width=7cm]{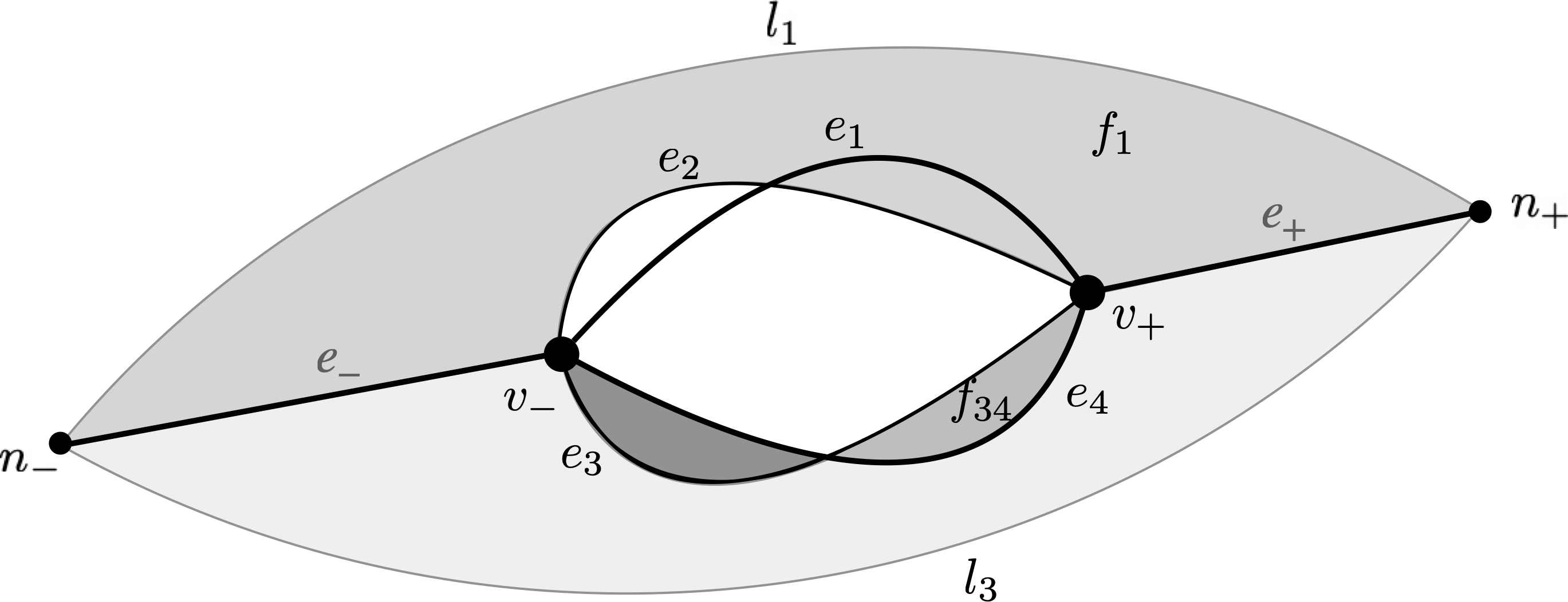}
    \caption{
    The two-complex $\cal C$ of the self-energy. All vertices and edges, but only one internal and two external faces are depicted.
    }
    \label{twocomplex}
\end{figure}

The dual of this cellular complex is a two-complex $\cal C$. It is formed by two pentavalent vertices $v_\pm$ joined by four internal edges $e_{a}$, with $a=1,2,3,4$. Each of the two vertices has one other edge $e_\pm$, ending on the boundary. The two-complex has four external faces $f_a$ that end on the boundary, and six internal faces $f_{ab}$, dual to the surfaces that separate the four internal tetrahedra. See Figure \ref{twocomplex}. The six faces $f_{ab}$, taken together, form the bubble we are interested in. 

\begin{figure}[h]
    \centering
    \includegraphics[width=7cm]{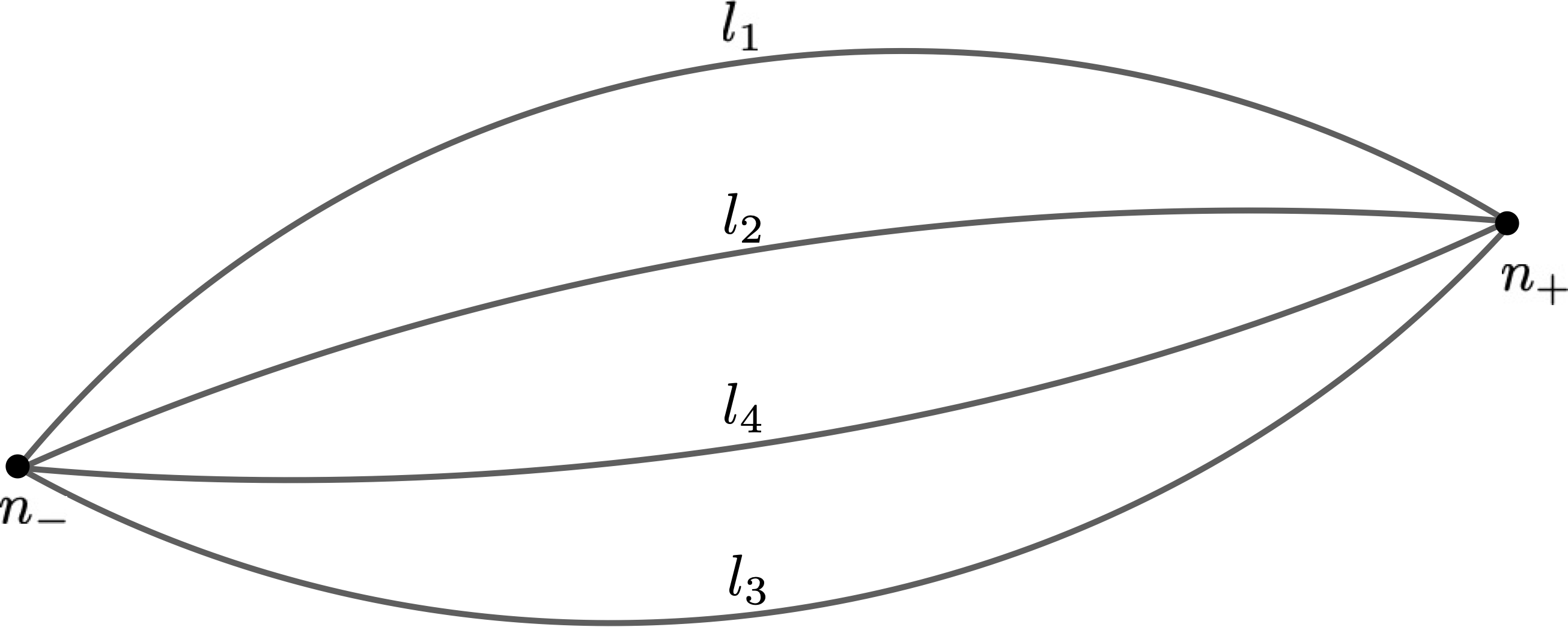}
    \caption{The boundary graph.}
    \label{fig:bubblediagram}
\end{figure}

The boundary of the two-complex $\cal C$ is the graph $\Gamma$ formed by the two 4-valent nodes $n_{\pm}$ where the edges $e_\pm$ end (dual to the initial and final quanta of space), joined by four links $l_a$ that bound the faces $f_a$ (see Figure \ref{fig:bubblediagram}). The kinematical LQG boundary Hilbert space is therefore $\mathcal{H}_{\Gamma} = L_2 \left[ SU(2)^4 / SU(2)^2 \right]_\Gamma$.  
The graph $\Gamma$ is represented in Figure \ref{fig:bubblediagram}. 

A basis in $\mathcal{H}_{\Gamma}$ is given by the spinnetwork states $|j_a,i_\pm\rangle$, where $j_a$ are the spins attached to the links $l_a$.   
In the two intertwiner spaces we choose the recoupling basis $i_\pm$ defined by pairing the links $j_1$ and $j_2$ at the node $n_{\pm}$, i.e. the basis 
that diagonalizes the modulus
square of the 
sum 
 of the $SU(2)$ generators in the $SU(2)$ representation $j_1$ and $j_2$ \cite{Bianchi:2011hp, Bianchi:2011ub,  Freidel:2010aq}. The label $i$ indicates the spin of the virtual link in this recoupling basis. 

We focus on the subspaces $\mathcal{H}_j\subset \mathcal{H}_\Gamma$ where the four spins are equal $j_a=j$, and denote the basis states as 
\begin{equation}
\label{eq:basis_states}
  |j, i_{\pm}\rangle  =  |j, i_+ \rangle \otimes  |j, i_{-}\rangle\ .
\end{equation}
We are interested in the EPRL amplitude \cite{Rovelli:2014ssa}
\be
W(j,i_\pm)=\langle W_{\cal C}|j, i_{\pm}\rangle
\ee
relative to
the two complex $\cal C$. To compute this amplitude numerically, we need an infrared cutoff to bound the sum over the spins associated to the internal faces. We denote this cutoff $K$, and we define the cutoff amplitude as
\be
W(j,i_\pm,K)=\langle W_{{\cal C},K}|j, i_{\pm}\rangle.
\ee
We are interested in studying the dependence of this amplitude on $K$.  

\section{BF amplitude} 
\label{sec:derivation_BF_amp}

As a first step, we consider the amplitude 
\be
W_{BF}(j,i_\pm)=\langle W^{BF}_{\cal C}|j, i_{\pm}\rangle
\ee
in the topological $SU(2)$ $BF$ theory, where the amplitude can be computed analytically \cite{Dona:2018infrared}.
The $BF$ self-energy amplitude can be written as  %
\begin{align}
W_{BF} & \left( j, i_\pm \right) =    \sum\limits_{ j_{ab}, i_a  }d_{\{j_{ab}\}} \, d_{\{i_{a}\}} \hspace{1mm} 
\prod_\pm \{15j\}_{j_a,j_{ab},i_\pm,i_a},
\label{eq:BF_simplicial_amplitude}
\end{align} 
where $d_{\{j_{ab}\}}=\prod\limits_{(a,b)}(2 j_{ab} +1)$ and $d_{\{i_{a}\}}=\prod\limits_{a}(2 i_{a} +1)$, with $a,b=1,...4$, $a \neq b$. The $BF$ vertex amplitude $ \{15j\}_{j_{AB},i_A}$ is the 15-j Wigner symbol defined contracting the five intertwiners $i_A$ with the ten spins $J_{AB}$, with $A,B=1,...5$, $A \neq B$. See the Appendix \ref{app:conventions} for notation and conventions. A graphical representation of this amplitude that will be useful later on is the following:
\begin{align}
\label{eq:BF_first_graph_exp}
W_{BF} & \left( j,i_\pm \right) = \sum\limits_{ j_{ab}, i_a }d_{\{j_{ab}\}} \, d_{\{i_{a}\}} \times   \\ &\nn 
  {\includegraphics[width=7cm]{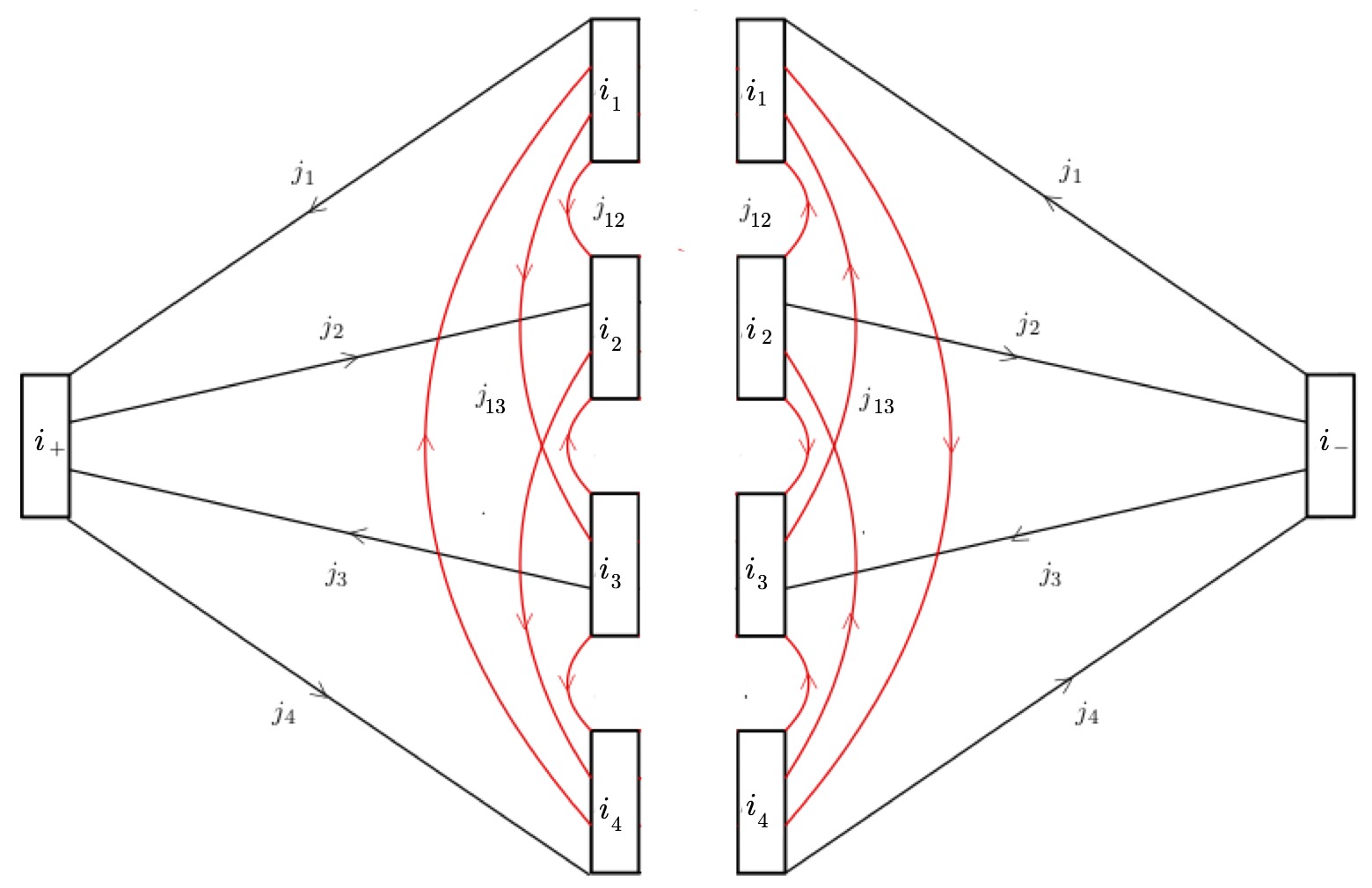}} . 
\end{align} 
Here for completeness we should  include in the expression for the amplitude the dimensional factor of the boundary faces and intertwiners, but this is a constant overall factor that can be ignored in the analysis since it does not affect the scaling of the divergence, that is what interests us. 
In \eqref{eq:BF_first_graph_exp}, the six internal faces $j_{ab}$ are highlighted in red. If we choose a recoupling basis for the internal intertwiners as follows
\begin{align}
W_{BF} & \left( j, i_\pm \right) = \sum\limits_{ j_{ab}, i_a }d_{\{j_{ab}\}} \, d_{\{i_{a}\}}  \times \\ &
{\includegraphics[width=7cm]{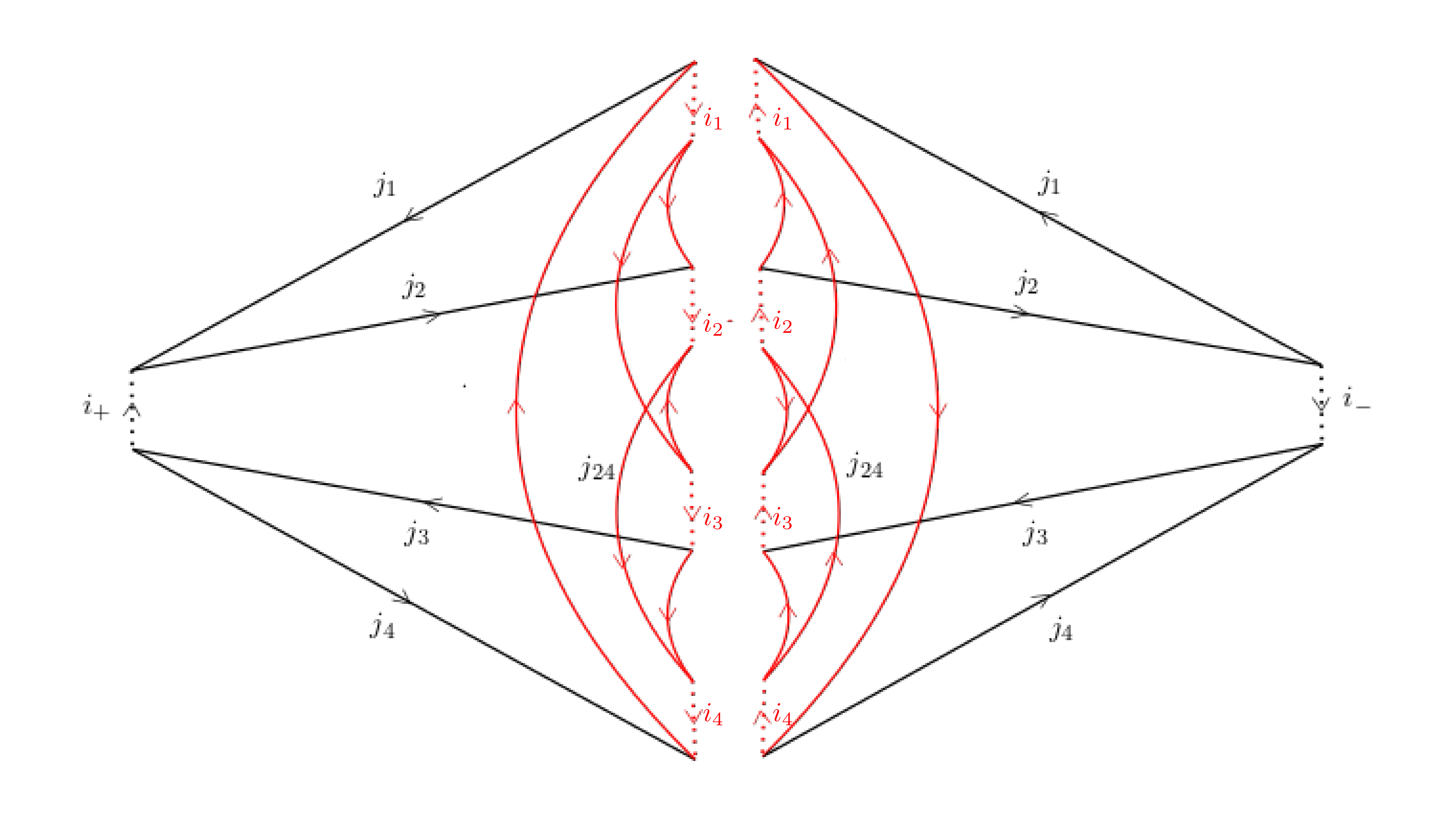}} \nn ,
\end{align} 
then this picture gives directly the expression of the two BF vertex amplitudes in terms of Wigner 3-j symbols: each intersection of three lines is one such symbols and the lines specify the contractions' scheme. 

\begin{figure}[!b]
    \centering
    \includegraphics[width=.9\linewidth]{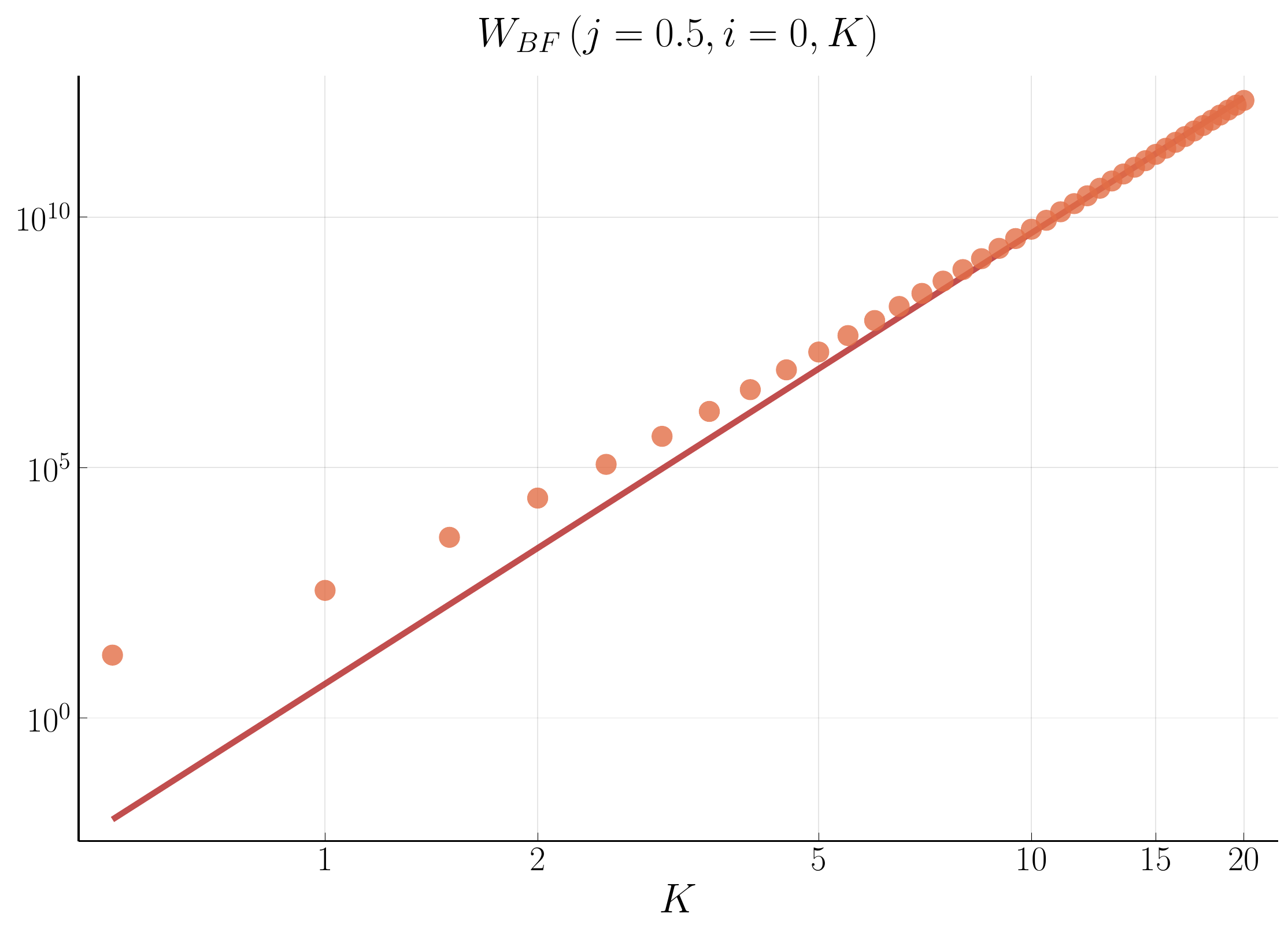}
    \caption{Log-log plot of the BF self-energy amplitude as a function of the cutoff $K$ on the internal spins $j_{ab}$. The continuous curve represents the function $4.8\!\cdot\!K^9$.\\ }
    \label{fig:bf_scaling}
\end{figure}

The multiple sum over the spins $j_{ab}$ and the virtual spins $i_a$ of the intertwiners are constrained by the Mandelstam identities that need to be satisfied for the Wigner symbols not to vanish. But it is clear from the last picture that the fixed values of the external spins $j_a$ do not prevent the internal spins $j_{ab}$ for growing arbitrarily large. 

\pagebreak
\noindent
This is the origin of the divergence. To study it numerically, we introduce a cutoff $j_{max}=K$ on the six internal faces $j_{ab}$, therefore bounding all the sums by $K$. That is, we define
\begin{align}
W_{BF} & \left( j, i_\pm ; K \right) = \sum_{\substack{ j_{ab} \leq K \\ i_a }}  d_{\{j_{ab}\}} \, d_{\{i_{a}\}} \hspace{1mm} 
\prod_\pm \{15j\}_{j_a,j_{ab},i_\pm,i_a} .
 \nn
\end{align}

We have computed this quantity numerically for a fixed value of the $j_a$ and $i_\pm$, for different values of $K$.  In the $BF$ case it is known analytically that the degree of divergence does not depend on $j_a$ and $i_\pm$.  The result is shown in Figure \ref{fig:bf_scaling}, where the continuous curve is $W_{BF} = 4.8 \cdot K^{9}$.
This result is consistent with the analytical result \cite{Dona:2018infrared}, which gives 
\begin{equation}
\label{eq:BF_mathematical_scaling_TOT}
W_{BF} \left( j, i_\pm ; K \right) \propto K^9.
\end{equation}
We briefly recall how to derive this result. Consider the simple case where all external spins are zero. The $BF$ amplitude can be expressed in Fourier transform as a multiple integral over $SU(2)$, given by one integration per link and one delta function per face. In this case we have  
\begin{equation*}
W_{BF} = \int \mathrm{d} g_{a} \prod_{a>b} \delta \left( g_{a} g_{b}^{-1}\right).
\end{equation*}
It is easy to see that one integration is redundant, and the other three can be used to eliminate three delta functions, leaving  
\begin{equation}
W_{BF} =  \delta \left( \mathds{1} \right)^3. 
\label{eq:BF_mathematical_scaling}
\end{equation}
To regularize this expression, one can write the delta function as a sum over characters $\chi^j$ of irreducible representations of $SU(2)$ and introduce an infrared cutoff $K$, as follows:
\begin{equation*}
\delta\left( U \right) = \lim_K \delta_K\left( U \right) = \lim_K \sum_j^K (2j+1) \chi^j \left(U\right) \ .
\end{equation*}
Then, easily, $\delta_K\left(\mathds{1} \right) \approx K^3$ and \eqref{eq:BF_mathematical_scaling_TOT} follows from \eqref{eq:BF_mathematical_scaling}.

\begin{figure}[!t]
    \centering
    \includegraphics[width=.9\linewidth]{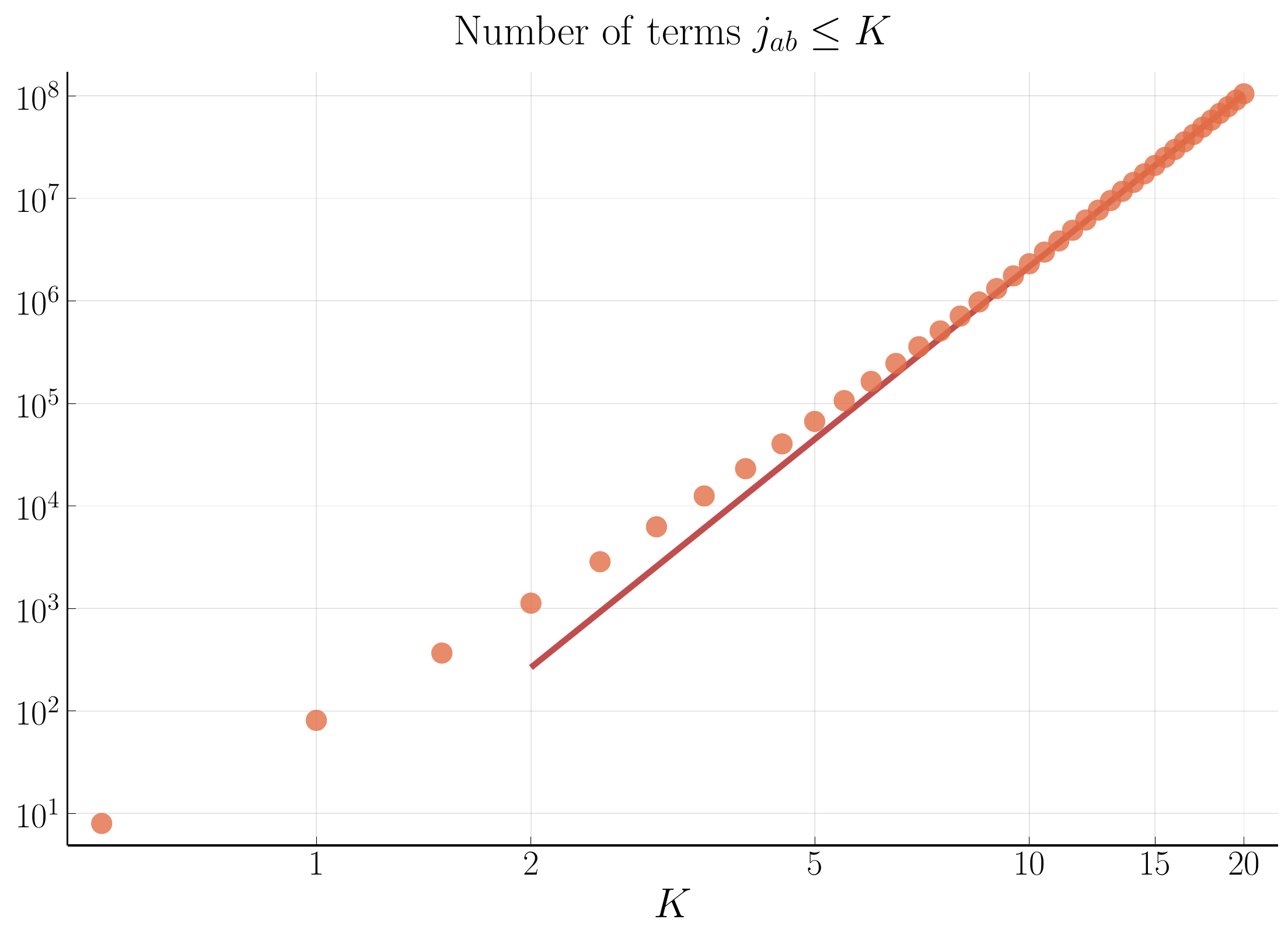}
    \caption{
    Log-log plot of the total number of different configurations of internal spins $j_{ab}$ to be summed over for increasing values of the cutoff $K$. The power-law fit is $\sim 1.7 \cdot K^{5.6}$.
    } 
    \label{fig:jab-number}
\end{figure}

The numerical analysis in Figure \ref{fig:bf_scaling} shows that the asymptotic scaling of the self-energy BF divergence \eqref{eq:BF_mathematical_scaling_TOT} is reached for very low values of the infrared cutoff $K$. The computation involved in producing Figure \ref{fig:bf_scaling} is nowadays trivial and can be easily performed on a laptop thanks to very efficient routines for computing Wigner (3,6,9)-j symbols \cite{johanssonFastAccurateEvaluation2015, Efficient_Storage_Scheme}. However, we note that the computational complexity grows for a large power of the cutoff index $K$, approximately equal to the number of internal faces of the foam. 
We plot in Figure \ref{fig:jab-number} the total number of combinations of internal spins that must be computed for each value of the cutoff $K$, till $K = 20$. We can see that the triangular inequalities reduce the complexity from $K^6$ to effectively $\sim K^{5.6}$. This still gives a large number of configurations to consider, and provides a considerable numerical challenge in the EPRL case, whose vertex amplitude is much more computationally complex than the BF one, as we show in the next section.
\section{EPRL}
\label{sec:From_BF_to_EPRL}
Let us now come to the quantum gravity theory.  
The EPRL vertex amplitude is defined using a restriction  
 of the unitary  
  irreducible  
  representations  
   in the  
   principal  
    series of $SL(2,\mathbb{C})$ \cite{Engle:2007wy,Engle:2007uq}. In order to evaluate it numerically, a more tractable formulation has been derived in \cite{Speziale2016} and discussed in \cite{Dona2018,article:Dona_etal_2019_numerical_study_lorentzian_EPRL_spinfoam_amplitude}. In this formulation, the vertex is expressed as a sum over $SU(2)$ 15-j symbols weighted by one ``booster function'' $B4$ per edge (see the Appendix \ref{app:booster} for explicit formulas). The amplitude of a vertex bounded by spins $j_f$ and intertwiners $i_e$ can thus be written as a sum
\be
\label{eq:first_EPRL_vertex}
V\!\!\left(j_f, \, i_e, \, \gamma \right) = \hspace{-1mm}\sum_{\substack{ l_f \geq j_f \\ k_e }}\hspace{-1mm} d_{\{k_{e}\}} \left( \prod_e
B_{4}(j_{f},l_{f};i_{e}, k_{e}; \gamma) \right) \{15j\}_{l_{f}, k_{e}} 
\ee
with virtual intertwiners $k_e$ and virtual spins $l_f \geq j_f$, $e=2...5$. One booster function must be replaced by the identity, corresponding to the suppression of a redundant $SL(2,C)$ integration\footnote{The 15-j symbol in \eqref{eq:first_EPRL_vertex} depends also on the 4 spins which label the gauge fixed edge, even if in the analytic notation it is not emphasized}. Notice that we keep the dependence on the Barbero-Immirzi parameter explicit. In this expression the infinite sum over $l_f$,  
called the ``shell expansion'', can be shown analytically to converge, but in view of the numerical calculation it is convenient to truncate it at a finite value defining 
\be
V\!\left(j_f, \, i_e , \gamma \right) = \lim_{\Delta l \to \infty} V\left(j_f, \, i_e , \gamma ; \Delta l \right)
\ee
where
\begin{align}
\label{eq:EPRL_vertex_amplitude}
&\hspace{-4mm}
V\!\!\left(j_f, \, i_e , \gamma ; \Delta l \right) =\\
& = 
\sum_{\substack{j_f\leq l_{f}\leq j_f + \Delta \\ k_e}}\hspace{-6mm}
d_{\{k_{e}\}} \left( \prod_e B_{4}(j_{f},l_{f};i_{e}, k_{e}; \gamma) \right)  \{15j\}_{l_{f}, k_{e}} \nn \\ \nn
& =   
\sum_{\substack{j_f\leq l_{f}\leq j_f + \Delta \\ k_e}}\hspace{-6mm}  
d_{\{k_{e}\}}
\raisebox{-2cm}{\includegraphics[width=5cm]{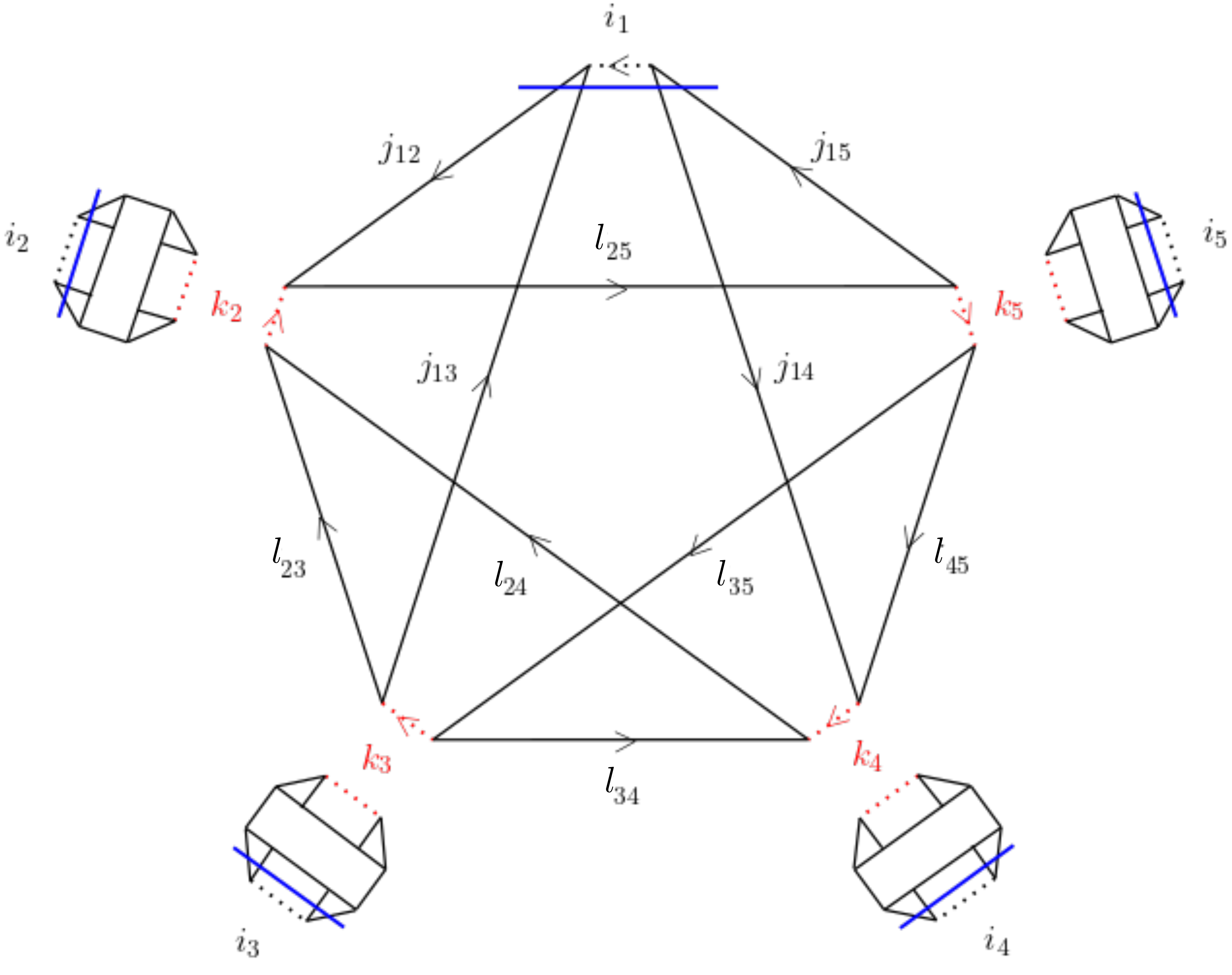}}. 
\end{align}
As before, we have neglected dimensional  
factors  
attached to  
the boundary 
 intertwiners  
$ i_e $ and to the boundary spins $j_f$. In the following, we will refer to the cutoff $\Delta l$  
as the  
number of \emph{shells}.
Even if the convergence of the amplitudes a function of $\Delta l$ is assured \cite{Dona2018, article:Dona_etal_2019_numerical_study_lorentzian_EPRL_spinfoam_amplitude}, it is not possible to have a unique prescription to set the optimal $\Delta l$ to get an acceptable convergence, since it depends on the details of data such as the face spins $j_f$ and the Barbero-Immirzi parameter. In addition, the convergence depends strongly on the structure of the 2-complex,  
and there is no known  
general procedure to  
estimate the error  
made in  
 truncating the sum  
  over the auxiliary spins. In the next Section we describe a technique to extrapolate  
  the limit  
   $\Delta l \rightarrow \infty$  
    for the self-energy. 

With the form of the vertex described above, we can write the self-energy amplitude explicitly, in a form suitable for the numerical analysis. This gives 
\begin{align}
\label{eq:EPRL_self_energy_amplitude}
& W\!\left( j, i_\pm, \gamma; K , \Delta l \right) =\hspace{-6mm} \sum\limits_{\substack{j_{ab}\leq K\\ j_{ab}\leq l_{ab} \leq  j_{ab}+\Delta l \\i_a ,  k_{q\pm}}} \hspace{-6mm}
d_{\{j_{ab}\}}d_{\{i_{a}\}} d_{\{k_{q\pm}\}} \times \nn \\ & 
 \includegraphics[width=8cm]{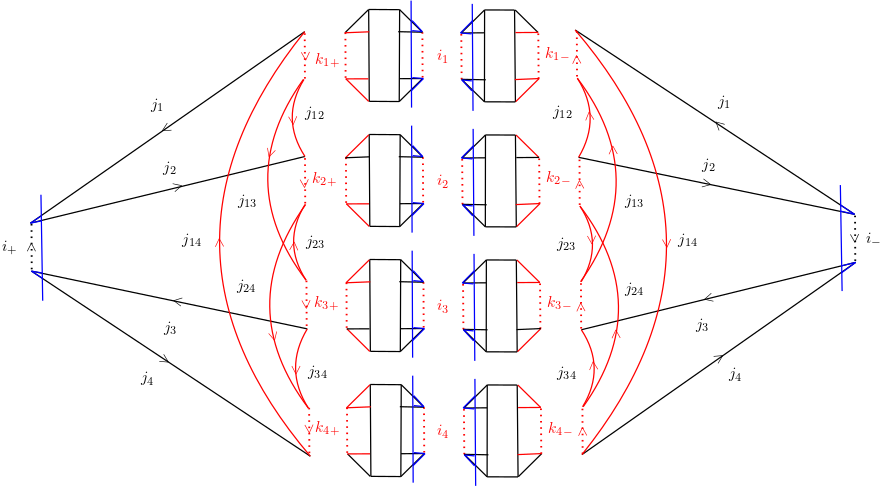},
\end{align}
with $q = 1...4$. Compared to the diagrammatic expressions used for the EPRL vertex \eqref{eq:EPRL_vertex_amplitude}, we have rearranged some links to emphasize the structure of the contractions between the vertices, which defines the spinfoam associated with the triangulation described in Section \ref{sec:self_energy_diagram}. 

The EPRL self-energy amplitude \eqref{eq:EPRL_self_energy_amplitude} is the main object of the analysis performed in this paper. The interested reader can derive the complete analytical expressions by comparing the graphical counterparts with the definitions given in Appendix \ref{app:conventions}.

When the two boundary  
intertwiners $i_{+}$ and $i_{-}$ have the  
same  
value  
$i$, the EPRL self-energy amplitude simplifies as
\begin{align}
\label{eq:EPRL_self_energy_amplitude_compact}
W\! \left( j, i, \gamma; K , \Delta l \right) = &  \sum_{\substack{ j_{ab} \leq K \\ i_a }} d_{\{j_{ab}\}}\, d_{\{i_{a}\}}
V(j, j_{ab}; i, i_a; \gamma ; \Delta l)^2
\end{align}
since the two vertex amplitudes entering the sum over the internal faces $j_{ab}$ are identical. Except for the coherent state (Appendix \ref{subsec:div_coherent_state}) and the boundary observable analysis (Appendix \ref{sec:Boundary_observables}), we shall always consider the two boundary intertwinwers $i_\pm$ to have the same value. In the following we use $i = (0,0)$. Result with different values are substantially identical, differing only by a slightly overall shift of the points, as we show in Appendix \ref{subsec:div_coherent_state}. 

Numerically, the vertex amplitude $V(j, j_{ab}; i, i_a; \gamma ; \Delta l)$ in \eqref{eq:EPRL_self_energy_amplitude_compact} can be efficiently computed using the recently developed library \texttt{sl2cfoam-next} \cite{Francesco_draft_new_code} given the list of boundary spins $(j_i = j, j_{ab})$ and the number of shells $\Delta l$. The squared sum over the internal intertwiner indices $i_a$ can be implemented by contracting the resulting \emph{vertex tensor} with itself. All the configurations of internal spins are automatically distributed by the library across the available cluster nodes, and each node parallelizes the computation of the various shells over the available local cpus. The particular code employed for this work is available at the repository \cite{self_energy_repo}. 

From \eqref{eq:EPRL_vertex_amplitude} it is clear that the computation of a single EPRL vertex amplitude is much more complex than the corresponding BF vertex amplitude with the same boundary spins and intertwiner indices. In fact, a single EPRL amplitude requires the computation of roughly $(\Delta l + 1)^6$ different BF amplitudes and $(\Delta l + 1)^3$ booster functions, \emph{for each} different set of intertwiners $i_a$ that bound the vertex (whose number is $\sim (2j + 1)^5$ if the boundary spins are of order $\sim j$). In the present case, a different EPRL vertex tensor (i.e.\ a bundle of amplitudes at fixed boundary spins with running boundary intertwiner indices) must be computed for each configuration of the internal spins $j_{ab}$, whose number as a function of the cutoff $K$ has been estimated in the previous Section (see Figure \ref{fig:jab-number}). From these simple estimates, we see that the numerical complexity of computing \eqref{eq:EPRL_self_energy_amplitude_compact}, using our methods, grows quickly when any of the parameters $K$, $\Delta l$ or $j$ increases, roughly with a power law between 5 to 6 powers of the increasing parameter. We provide in the next Section some data for the resources employed in a computation of \eqref{eq:EPRL_self_energy_amplitude_compact} up to $K = 10$ and $\Delta l = 20$.
\pagebreak

\section{Divergence Analysis}
\label{sec:Div_analysis}
We have studied the amplitude \eqref{eq:EPRL_self_energy_amplitude_compact} numerically and considered the following questions:
\begin{enumerate}
    \item\label{Question_1} What is the asymptotic scaling of the EPRL divergence? 
    \item\label{Question_2} What is the dependence on the Barbero-Immirzi parameter? 
    \item\label{Question_3} At fixed boundary spins and Barbero-Immirzi parameter, does the scaling change using boundary spinnetwork states or intrinsic coherent states? 
    \item\label{Question_4} What are the configurations which contribute most to the divergence itself?
\end{enumerate}
The first two questions are addressed in the main text, the third in Appendix \ref{subsec:div_coherent_state} and the fourth in Appendix \ref{app:internal_contributions}. The main difficulty towards answering these questions is the limit $\Delta l\to \infty$ that cannot be taken numerically. To address this problem, we have relied on a property of the dependence of the amplitude on  $\Delta l$ at fixed $K$, that has emerged from the numerical analysis itself. 
\begin{figure*}[!hbtp]
    \centering
    \begin{subfigure}[b]{.49\textwidth}
        \includegraphics[width=\linewidth]{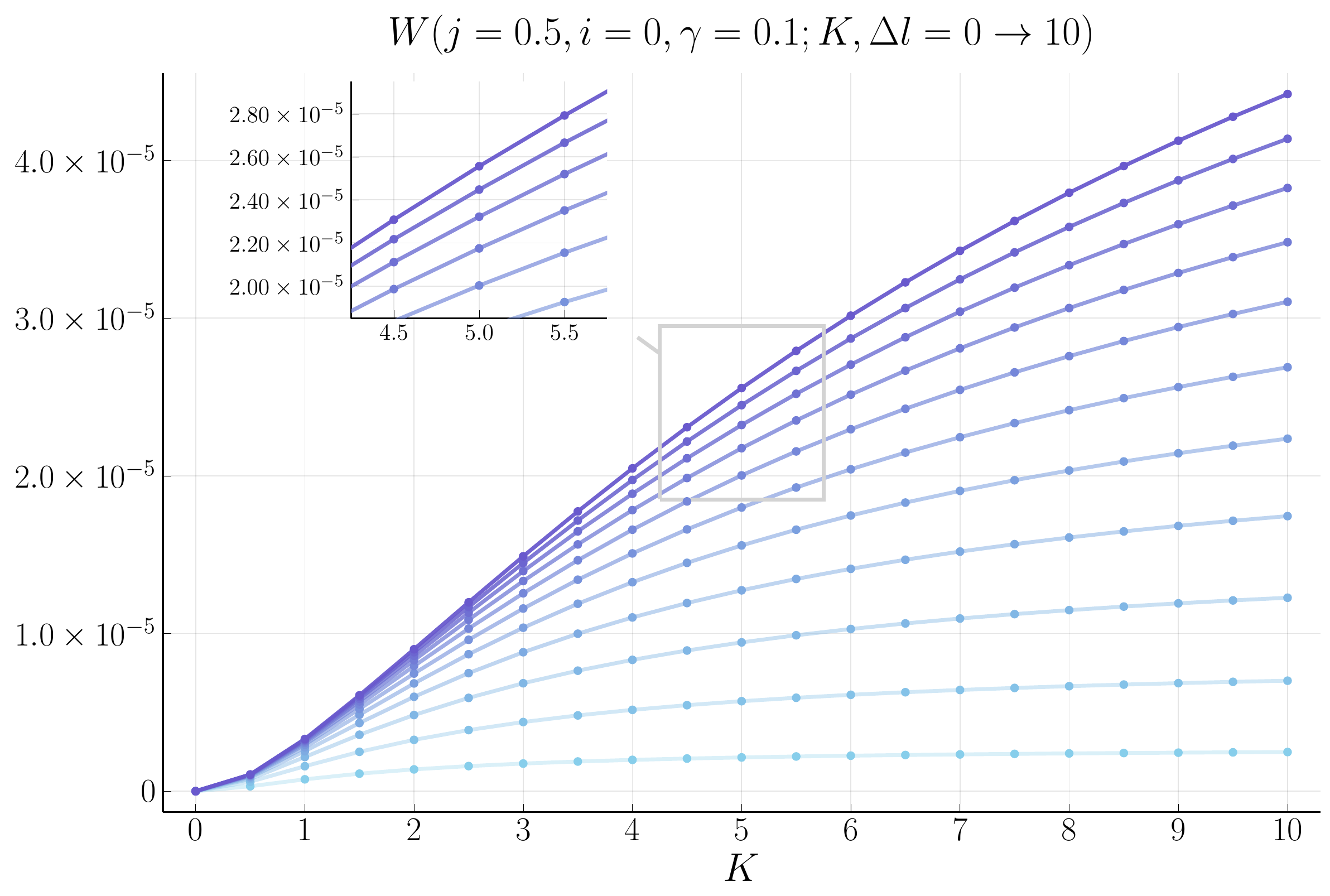}
    \end{subfigure}  
    \begin{subfigure}[b]{.49\textwidth}
        \includegraphics[width=\linewidth]{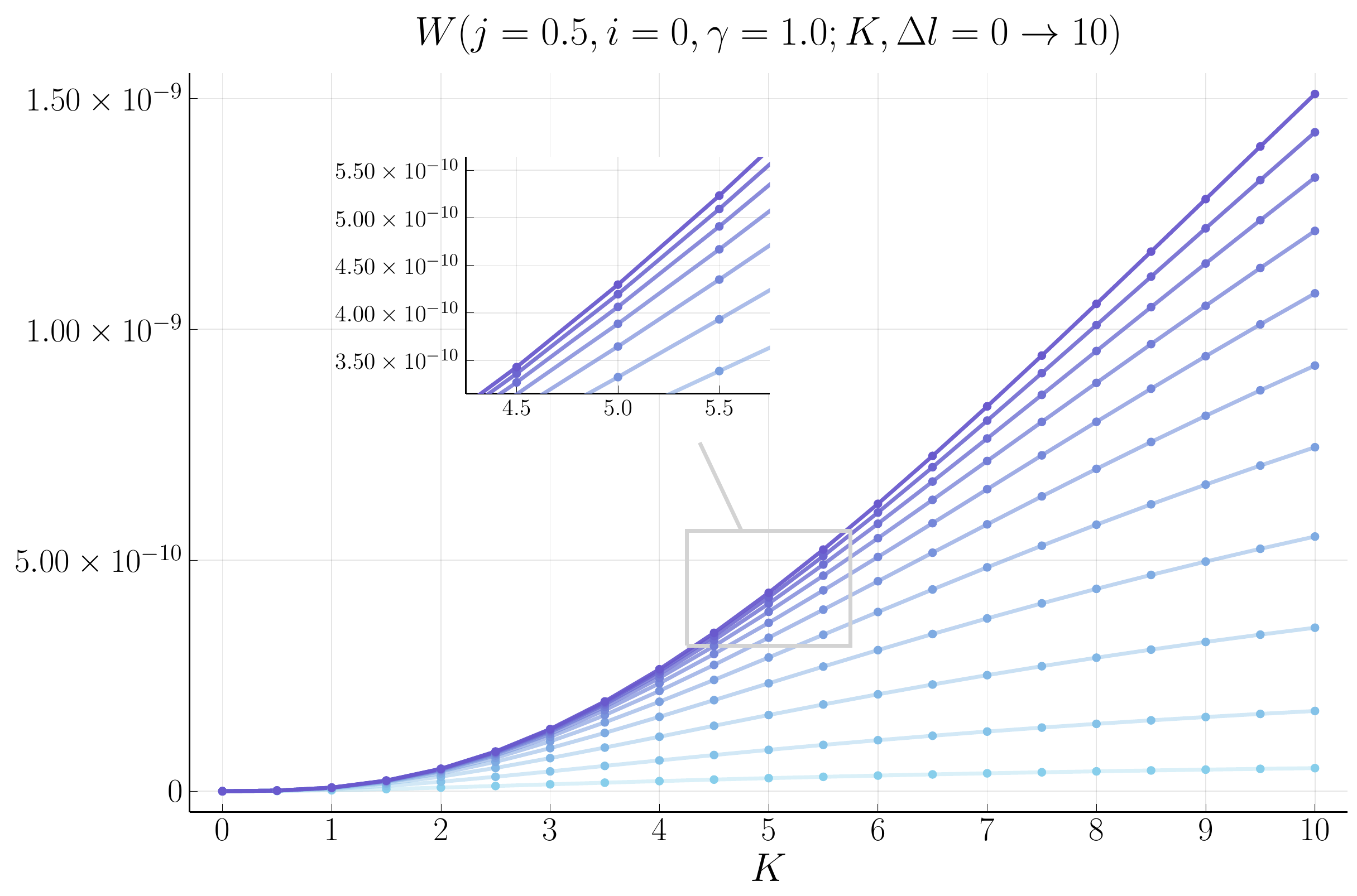}
    \end{subfigure}  

    \caption{ 
    \label{fig:shells-2gs}
    The EPRL divergence \eqref{eq:EPRL_self_energy_amplitude_compact} as a function of the cutoff parameter $K$, for a number of shells $\Delta l = 0$ to $10$. Darker colors correspond to larger $\Delta l$. A portion of the plot around $K = 5$ is highlighted to show the convergence properties of the shells approximation. \emph{Left:} $\gamma = 0.1$. \emph{Right:} $\gamma = 1.0$.
    }
\end{figure*}

\subsubsection*{Extrapolation scheme}

We show in Figure \ref{fig:shells-2gs} the amplitude \eqref{eq:EPRL_self_energy_amplitude_compact} as a function of the cutoff $K$, for various values of the number of shells $\Delta l$ and for $\gamma = 0.1$ and $\gamma = 1.0$. From a simple qualitative analysis we infer that the convergence of the shell approximation is faster for a reduced bulk spins cutoff, while it becomes slower as $K$ increases. The highest curve is a lower bound to the amplitude as a function of $K$, which is recovered in the limit $ \Delta l \rightarrow \infty$. Convergence in the number of shells appears to depend on $\gamma$, with the case $\gamma = 1.0$ converging faster.

\begin{figure}[!b]
\centering
\includegraphics[width=\linewidth]{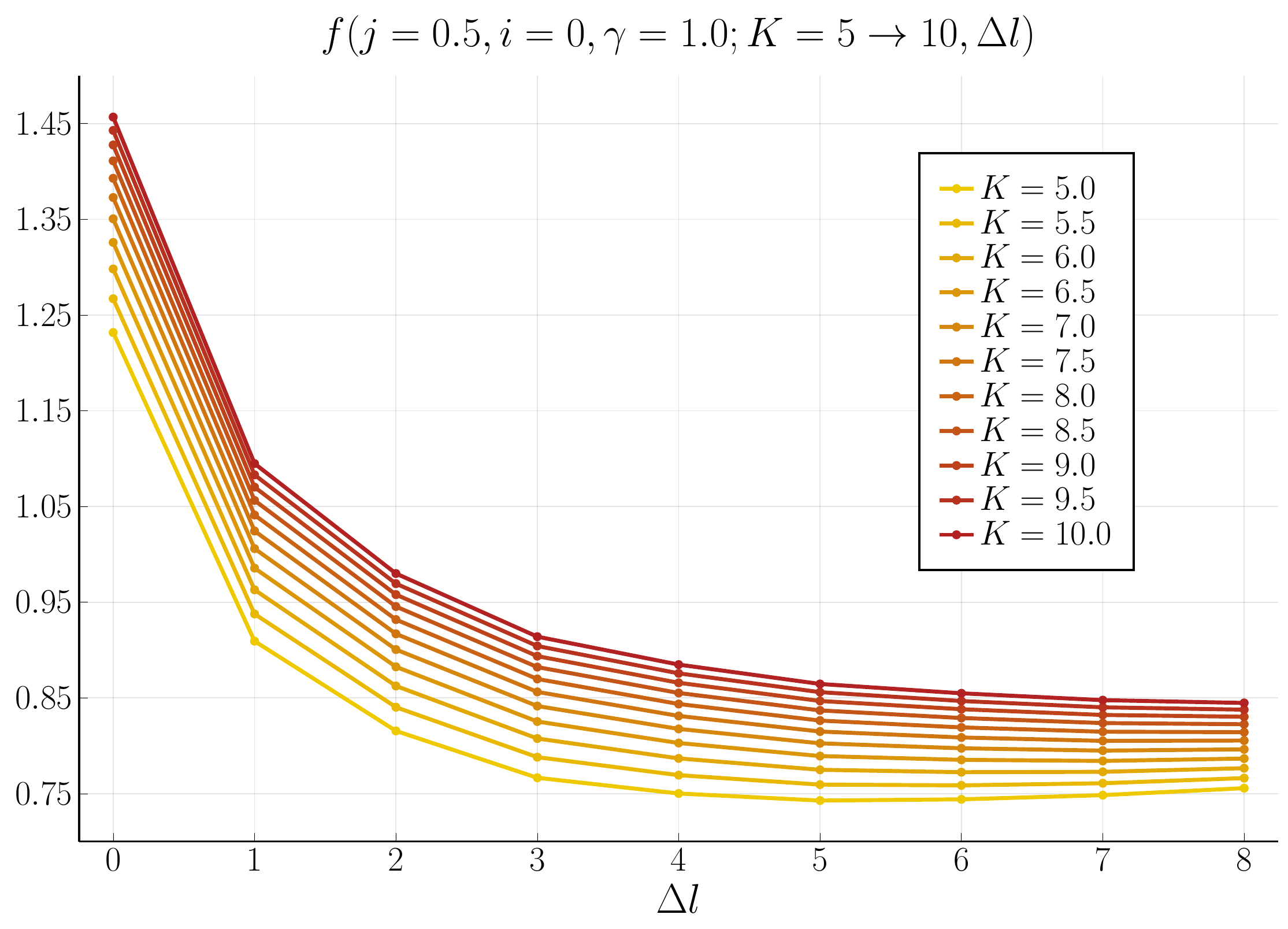}
\caption{\label{fig:ratios-g10} Function \eqref{eq:f_function} for $\gamma = 1.0$. The curves appear to tend to a constant value $c_{K,\gamma}$. }
\end{figure}

The convergence of \eqref{eq:EPRL_self_energy_amplitude_compact} in the parameter $\Delta l$ can be extrapolated from the data reported in Figure \ref{fig:shells-2gs}. To this aim, we have computed the ratios  
of the  
differences  
between  
adjacent  
curves of Figure \ref{fig:shells-2gs}, at 
 fixed 
  $K$, as a  
  function  
  of $\Delta l$. That 
   is, we  
   have  
   studied  
   the function:
\begin{equation}
\label{eq:f_function}
f(\gamma, K, \Delta l) \equiv \frac{W(\gamma, K, \Delta l + 2) - W(\gamma, K, \Delta l + 1)}{W(\gamma, K, \Delta l + 1) - W(\gamma, K, \Delta l)} \ ,  
\end{equation}
for each fixed value of  $j$ and $i$. 
The result is shown in Figure \ref{fig:ratios-g10} for $K\geq 5$ and $\gamma = 1.0$.
It appears from Figure \ref{fig:ratios-g10} that the function $f(\gamma, K, \Delta l)$ tends to a constant for larger enough values of $K$, as a function of $\Delta l$ (however, see also the more complete analysis for the case $\gamma = 0.1$, that is studied in a following section). Since \eqref{eq:f_function} is the ratio between successive terms of a series, this means that the sum over the parameter $\Delta l$ can be well approximated by a geometric series when $K$ is large. Defining
\begin{equation}
\label{eq:c_coefficient}
c_{K,\gamma} \equiv f\left(\gamma, K, N-2 \right)
\end{equation}
where $N$ is the largest $\Delta l$ parameter that has been reached numerically, our assumption implies for $\Delta l \geq N$
\begin{equation}
\label{eq:pipeline_property}
W(\gamma, K, \Delta l + 1) - W(\gamma, K, \Delta l) \approx \left( c_{K, \gamma} \right)^{\Delta l}\ .  
\end{equation}
We thus approximate the divergence in the limit $\Delta l \to \infty$ as
\begin{align}
\label{eq:extrapolation_equation}
W(\gamma, K) &\equiv \lim_{\Delta l \to \infty} W(\gamma, K, \Delta l) 
\\ \nn 
&\!\!\!\!\!\!\!\!\!\!\!\!\!\!\!\!
\approx W(\gamma, K, N-1) + \frac{W(\gamma, K, N) - W(\gamma, K, N-1)}{1 - c_{K, \gamma}}, 
\end{align}
where we used the elementary limit of the geometric series. We note that the approximation improves when \emph{(i)} more shells can be computed numerically, i.e. the index $N$ increases, and \emph{(ii)} the ratios \eqref{eq:f_function} are closer to true constants in $\Delta l$. We applied this extrapolation scheme to the amplitudes computed numerically for different values of $\gamma$. In particular, in the case $\gamma = 0.1$, where the convergence in $\Delta l$ is the slowest, hence the approximation is the least accurate, we pushed the $\Delta l$ parameter to very large values in order to study the accuracy of our extrapolation scheme and obtain very precise results for a physically relevant case.

\subsubsection*{ Case $\gamma = 0.1$}
\begin{figure*}[!hbtp]
    \centering
    \begin{subfigure}[b]{.49\textwidth}
        \includegraphics[width=\linewidth]{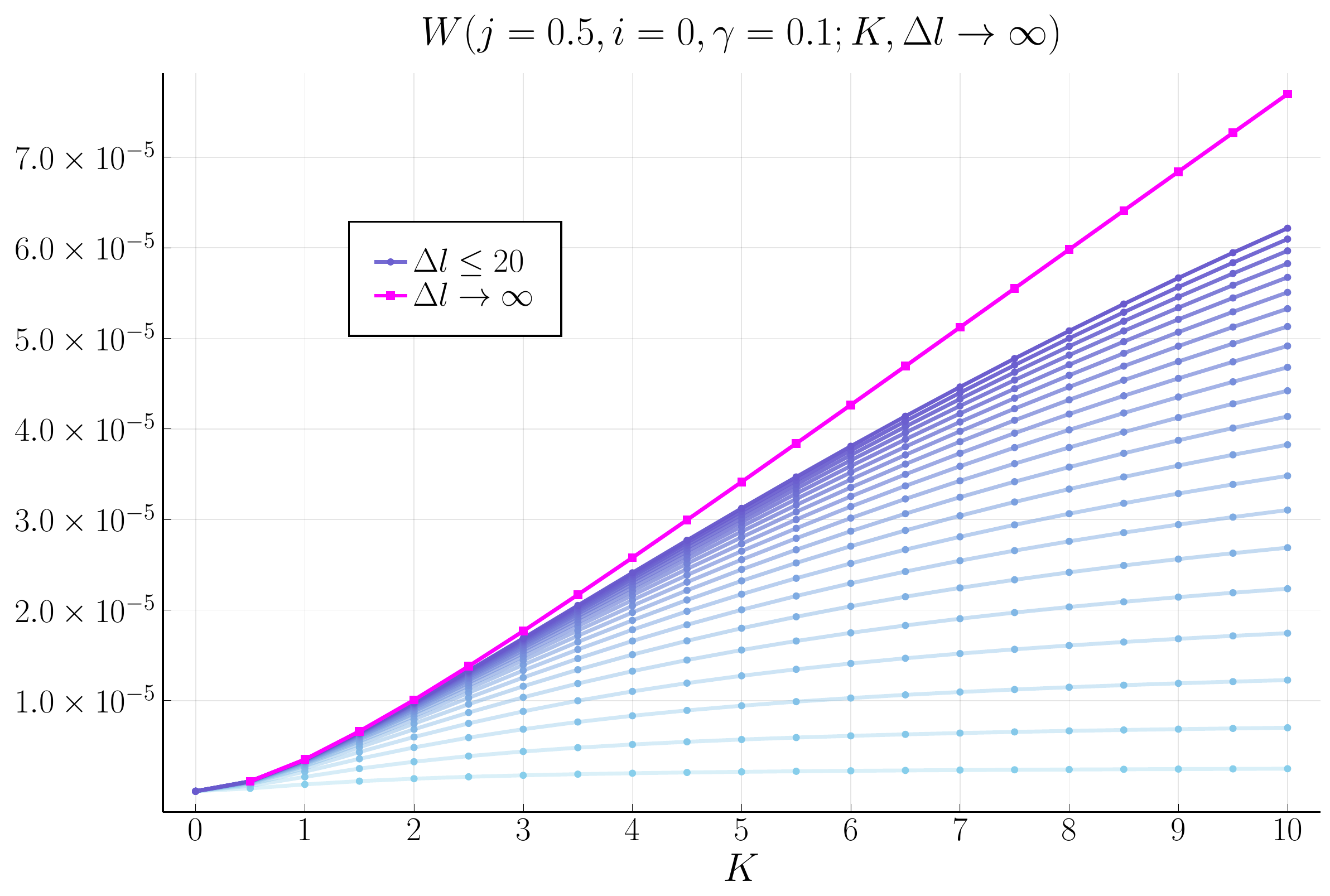}
    \end{subfigure}  
    \begin{subfigure}[b]{.49\textwidth}
        \includegraphics[width=\linewidth]{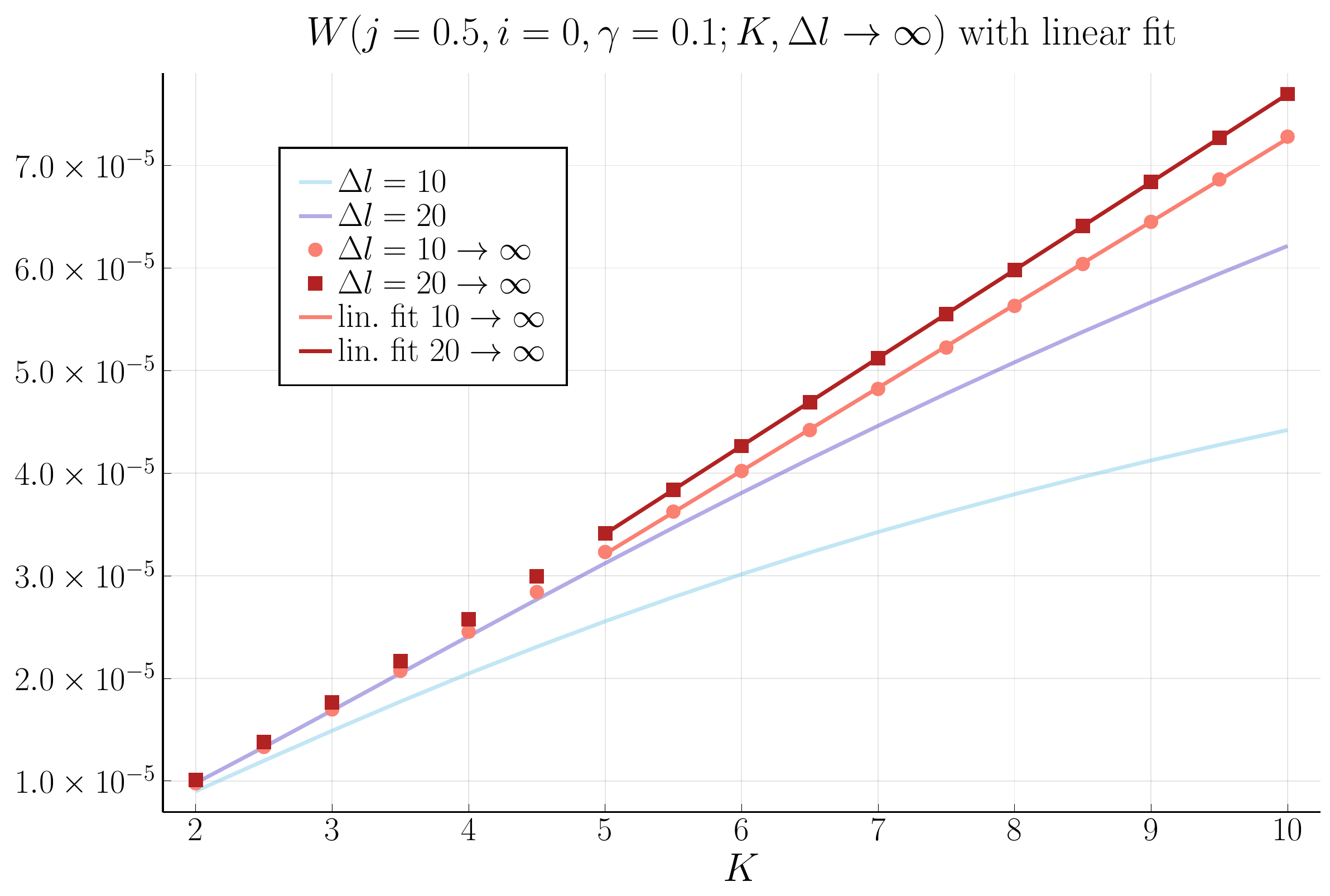}
    \end{subfigure}  

    \caption{ 
    \label{fig:fits_g01}
    The complete analysis for the case $\gamma = 0.1$ till $N = \Delta l_\mathrm{max} = 20$. \emph{Left:} All the curves at various $\Delta l$ with the extrapolation from $\Delta l = 20$. \emph{Right:} The extrapolations from $\Delta l = 10$ and $\Delta l = 20$ with the corresponding linear fits. The slopes of the fits are $8.1 \cdot 10^{-6}$ and $8.6 \cdot 10^{-6}$, respectively. The standard errors of the fits, obtained with Julia's \texttt{LsqFit} package, are $2.3 \cdot 10^{-8}$ and $4.6 \cdot 10^{-9}$, respectively.
    }
\end{figure*}
\begin{figure}[b]
\centering
\includegraphics[width=\linewidth]{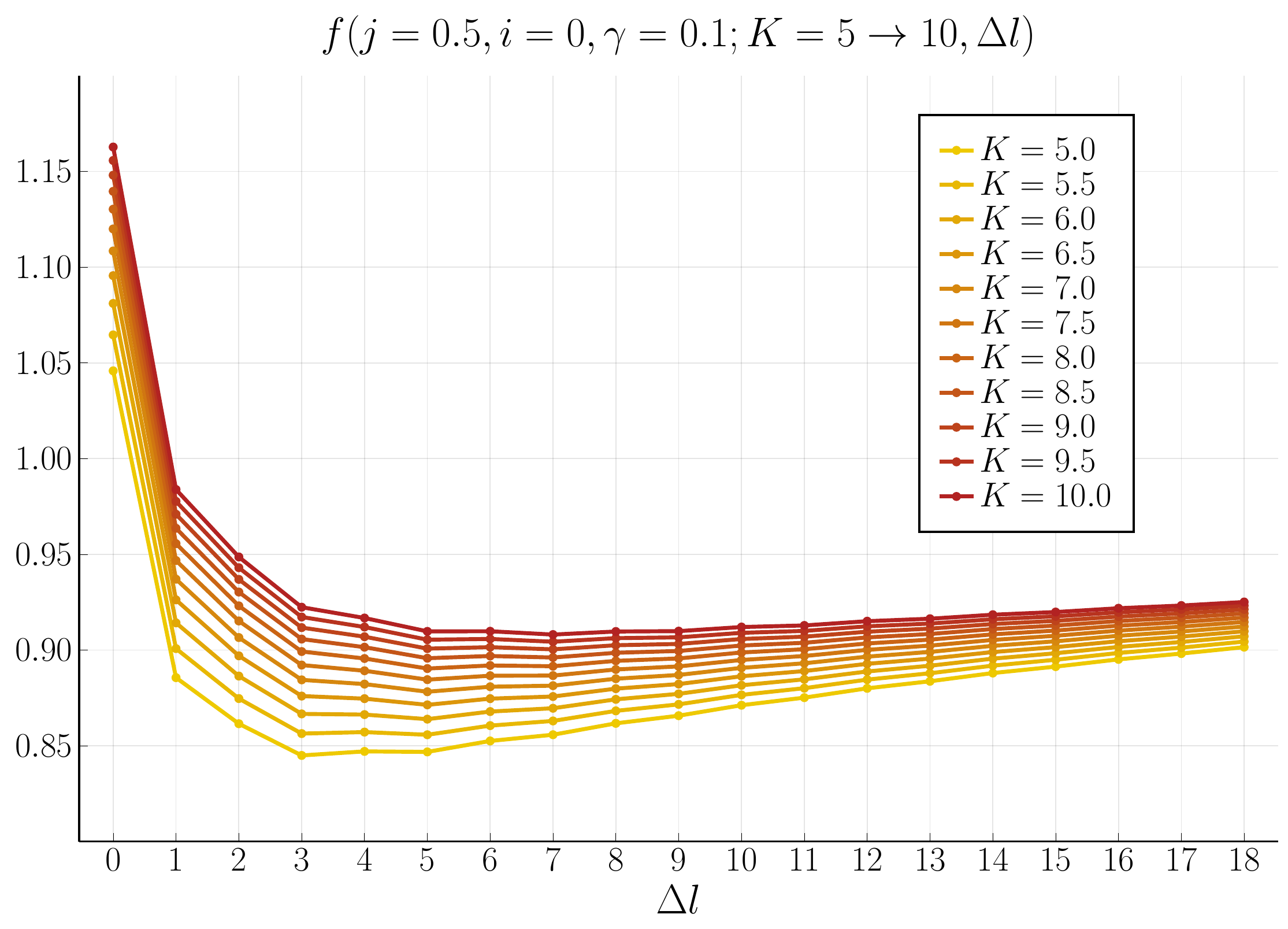}
\caption{
\label{fig:ratios-g01-long} 
Function \eqref{eq:f_function} for $\gamma = 0.1$ plotted till $N = \Delta l_\mathrm{max} = 20$. It is evident that the approximation $f \approx \mathrm{const}$ for $\Delta l > N$ provides a lower bound to the exact limit $\Delta l \to \infty$.
}
\end{figure}
In Figure \ref{fig:fits_g01} we show the amplitude \eqref{eq:EPRL_self_energy_amplitude_compact} for the case $\gamma = 0.1$, with the computed data for $\Delta l \leq 20$ and the extrapolated curves from $N = 10$ and from $N = 20$. From these plots we can verify that the approximation gives results that are already accurate starting with low values of $N$, since the difference between the two extrapolations is small, around 5\% for $K = 10$. Plotting the ratios \eqref{eq:f_function} till very large values of $\Delta l$, as we did in Figure \ref{fig:ratios-g01-long}, we can see that for the case $\gamma = 0.1$ the approximation $f(\gamma, K, \Delta l) \approx \mathrm{const}$ is more accurate for larger values of $K$ and less accurate for smaller cutoffs. This is not an issue since convergence in $\Delta l$ is reached sooner for small cutoffs (since the ratios are smaller, the series converges faster). Also, Figure \ref{fig:ratios-g01-long} shows that the extrapolation is actually a lower bound to the true amplitude, although very close to exact result.

Importantly, the qualitative behavior of the divergence, either extrapolated from $N = 10$ or $N = 20$, is clearly linear. In Figure \ref{fig:fits_g01} we show the fit of the extrapolated curves with straight lines. It is clear that the extrapolated points fall exactly on a straight line for $K \gtrsim 5$ in both cases, with the better approximation at $N = 20$ being also closer to linear behavior than the one at $N = 10$. We conclude in this case that the overall divergence is very close to linear in the cutoff parameter $K$. 

A few comments are in order. First, our results are in accordance with the rough analytical estimate of Riello \cite{Riello:2013bzw}. It is important to notice that his main analysis focuses on what Riello calls the \emph{non-degenerate} sector of the divergence, and disregards the contributions from \emph{degenerate} configurations when the 4-simplices dual to the two spinfoam vertices $v_\pm$ have non-maximal dimension. He found that this sector contributes with a leading factor of $\log K$ to the divergence. However, in Appendix C of \cite{Riello:2013bzw} the remaining \emph{degenerate} sector is estimated to contribute with a dominant factor of $K$ to the overall divergence, which is what our numerical analysis shows. Our result also clearly shows that the interference effects neglected in \cite{Dona:2018infrared} result in a strong suppression of the divergence, of approximately 8 powers of $K$.

Second, our numerical analysis does not completely excludes the possibility that the behavior for very large values of $K$ is different from what is inferred in the regime $K \lesssim 10$. There are however multiple arguments supporting our conclusion: \emph{(i)} our findings are compatible with past analytical estimates \cite{Riello:2013bzw}; \emph{(ii)} from Figure \ref{fig:fits_g01} we see that improving the approximation also improves the matching with the linear fit, which appears perfect for as much as about 12 data points; \emph{(iii)} we can compare the EPRL divergence with the BF case, where the asymptotic power-law divergence is already manifest for spins of order $\sim 10$ (see Figure \ref{fig:bf_scaling}) and analytically it is seen to be independent from the details of the chosen boundary data; \emph{(iv)} recent numerical investigations \cite{Francesco_draft_new_code, asanteEffectiveSpinFoam2020} strongly support the hypothesis that the semiclassical (i.e.\ asymptotic) regime of theory is reached for relatively small spins if the Barbero-Immirzi parameter $\gamma$ is small (e.g.\ around $0.1$). This appears to be related to the slower frequency of asymptotic oscillations of the amplitude whenever $\gamma$ is small \cite{asanteEffectiveSpinFoam2020}. Hence, we can assume that we are looking at the asymptotic divergence for $K \gtrsim 5$, in this case. This is also supported by our analysis for larger values of $\gamma$, which do not appear to converge to a fixed ($\gamma$-independent) asymptotic behavior in the considered range of the cutoff, as we show in the next paragraph. 

\begin{figure*}[!tbp]
    \centering
    \begin{subfigure}[b]{.3\textwidth}
        \includegraphics[width=\linewidth]{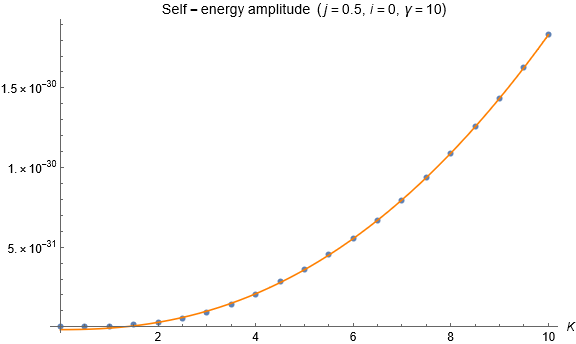}
    \end{subfigure}  
    \begin{subfigure}[b]{.3\textwidth}
        \includegraphics[width=\linewidth]{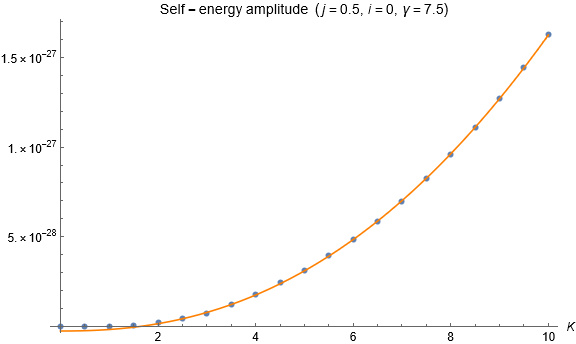}
    \end{subfigure}  
    \begin{subfigure}[b]{.3\textwidth}
        \includegraphics[width=\linewidth]{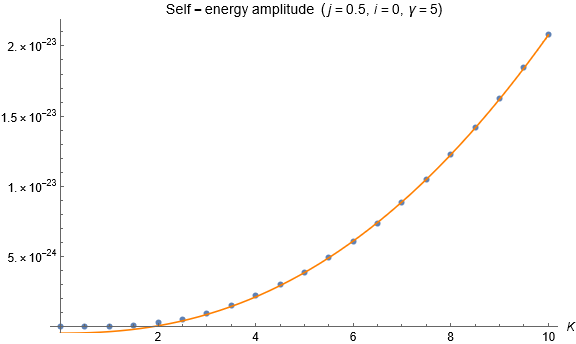}
    \end{subfigure}   
  \begin{subfigure}[b]{.3\textwidth}
        \includegraphics[width=\linewidth]{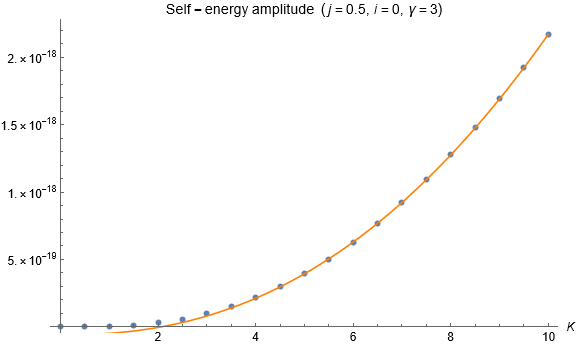}
    \end{subfigure}        
  \begin{subfigure}[b]{.3\textwidth}
        \includegraphics[width=\linewidth]{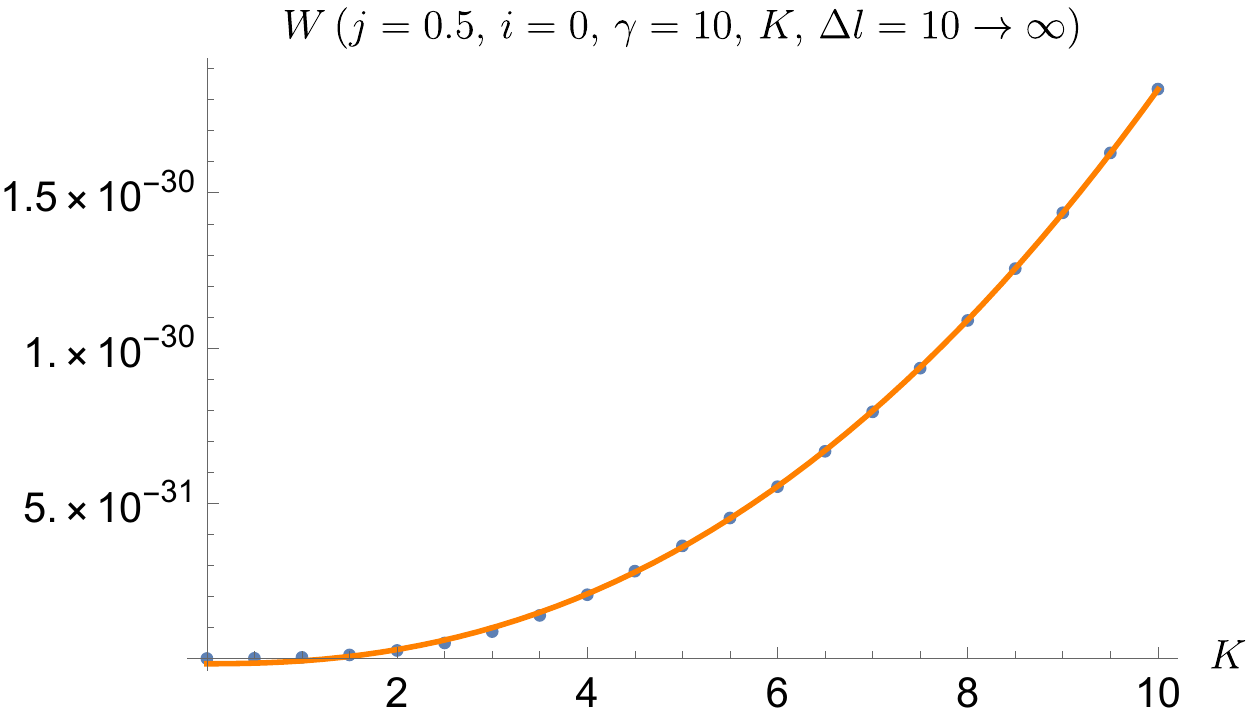}
    \end{subfigure}     
  \begin{subfigure}[b]{.3\textwidth}
        \includegraphics[width=\linewidth]{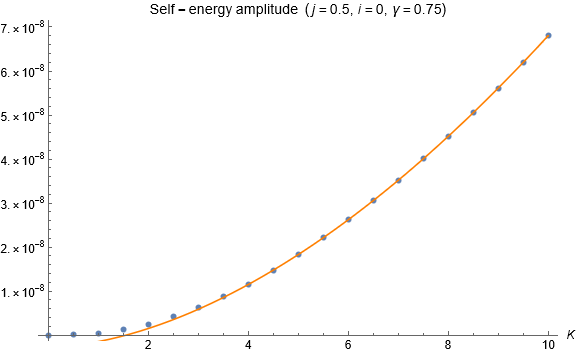}
    \end{subfigure}   
     \begin{subfigure}[b]{.3\textwidth}
        \includegraphics[width=\linewidth]{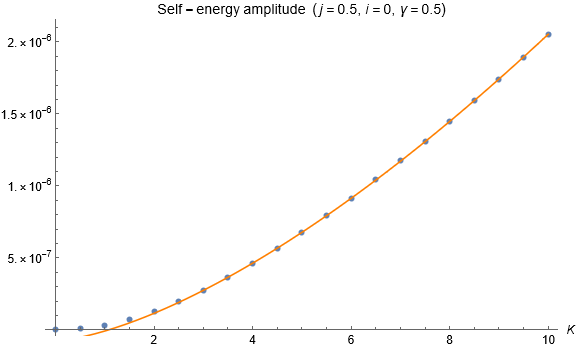}
    \end{subfigure}
          \begin{subfigure}[b]{.3\textwidth}
        \includegraphics[width=\linewidth]{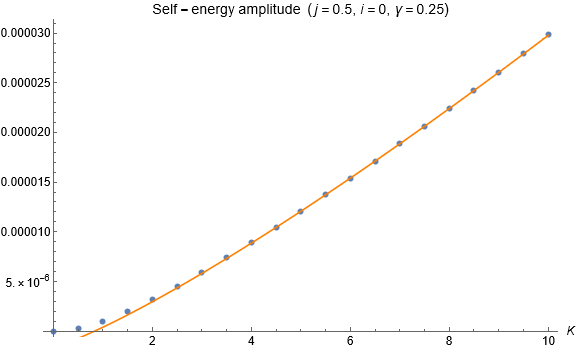}
    \end{subfigure}    
          \begin{subfigure}[b]{.3\textwidth}
        \includegraphics[width=\linewidth]{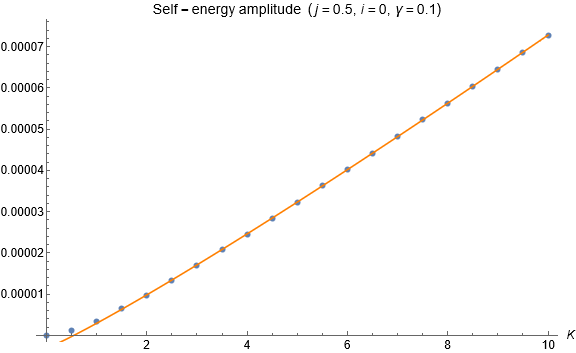}
    \end{subfigure}
    \caption{ 
    \label{fig:increasingscaling}
    Fit of the amplitude $W(\gamma, K)$ extrapolated via equation \eqref{eq:extrapolation_equation} with $N=10$ for decreasing values of the Barbero-Immirzi parameter $\gamma \neq 0.1$ and with $N=20$ for $\gamma = 0.1$. The fit is made cutting the values of $W(\gamma, K)$ for $K \leq 4$ and  superimposing the resulting curve with all the extrapolated points. 
    }
\end{figure*}

As a final comment for the $\gamma = 0.1$ case, we report some technical details about the hardware resources that were employed. The computation of the divergence at $K \leq 10$ and $\Delta l = 20$ ran on 640 cpus for $\sim 60$ hours, for a total of $\sim 40000$ cpu-hours. The whole computation at $\gamma = 0.1$ from $\Delta l = 0$ to $20$ took about $125000$ cpu-hours. Recall that the computational complexity of the EPRL model scales as $(\Delta l + 1) ^ 6$, thus the plots at largest number of shells took most of the allocated cpu-time. Symmetries in the internal spins $j_{ab}$ peculiar to this case ($j = 1/2, i_\pm = 0$) have been implemented in the code to reduce the computational time of about a factor $\sim 6$.

\subsubsection*{Case $\mathbf \gamma \gtrsim 0.1$}

The extrapolation method outlined above allowed us to investigate also larger values of the Barbero-Immirzi parameter $\gamma$ without replicating the already substantial computational effort required for the case $\gamma = 0.1$. We have studied the scaling of the asymptotic divergence for $9$ values of $\gamma$, ranging between $0.1$ and $10$. The result is shown in Figure \ref{fig:increasingscaling}, for which we used $N=10$ in equation \eqref{eq:extrapolation_equation}. Notice that the extrapolation scheme is more effective in the cases $\gamma > 0.1$ than in the case $\gamma = 0.1$, as the curves of the ratios given by the function \eqref{eq:f_function} become closer to constants as $\gamma$ increases (compare Figure \ref{fig:ratios-g10} with Figure \ref{fig:ratios-g01-long}).
The data of Figure \ref{fig:increasingscaling} are fitted with a function:
\begin{equation}
W(\gamma, 
 K) = a  
 + b K^{c}
\label{eq:fitting_form}
\end{equation}
where $a,b,c$ are  
real  
coefficients. The best fit for the values of these parameters are shown in Table \ref{tbl:coefficient_values}. 
\begin{center}
\begin{table}[!htbp]
\caption{Fit coefficients table}
\begin{tabular}{ |p{1cm}||p{2cm}|p{2cm}|p{1cm}|  }
 \hline
 \multicolumn{4}{|c|}{ $W(\gamma, K) = a + b K^{c}$ $ \hspace{2mm}  (j = 0.5, i = 0)$, $K \in [0, 10]$} \\
 \hline
$\gamma$ & $a$ & $b$ & $c$ \\
 \hline
 $10$   & $-1.7 \cdot 10^{-32}$ & $9.2 \cdot 10^{-33}$ & 2.3  \\
 $7.5$   & $-2.2 \cdot 10^{-29}$ & $8.2 \cdot 10^{-30}$ & 2.3  \\ 
 $5$   & $-4.3 \cdot 10^{-25}$ & $1.1 \cdot 10^{-25}$ & 2.3  \\
 $3$   & $-5.7 \cdot 10^{-20}$ & $1.1 \cdot 10^{-20}$ & 2.3  \\
 $1$ & $-8.6 \cdot 10^{-11}$ & $2.6 \cdot 10^{-11}$ & 1.9 \\
 $0.75$ & $-1.7 \cdot 10^{-9}$ & $1.1 \cdot 10^{-9}$  & 1.8 \\
 $0.5$ &  $-7.6 \cdot 10^{-8}$  &  $6.7 \cdot 10^{-8}$   & 1.5 \\
 $0.25$ & $-1.5 \cdot 10^{-6}$ & $2.0 \cdot 10^{-6}$ & 1.2 \\
 $0.1$  & $-8.8 \cdot 10^{-6}$    & $8.6 \cdot 10^{-6}$ &  1  \\
 \hline
\end{tabular}
\label{tbl:coefficient_values}
\end{table}
\end{center}
Table \ref{tbl:coefficient_values} shows that the amplitude is suppressed for large $\gamma$. The divergence is increasingly well fitted by an approximately linear scaling as the Barbero-Immirzi parameter decreases, while for $\gamma \gtrsim 1$ the curve is approximated by a  
quadratic  
function  
in the  
range  
$K \in \left[0,10 \right]$. We interpret this apparent dependence of the scaling on $\gamma$ as a sign that the asymptotic or semiclassical regime is not reached for $K \leq 10$ when $\gamma \gtrsim 1$. A possible educated guess is that the non-linear behavior for large $\gamma$ could correspond to the initial non-linear part of the curve seen also at small $\gamma$, namely that the dependence could become linear for any $\gamma$ at sufficiently high $K$. This intriguing hypothesis could have interesting consequences for a renormalization procedure. We leave the accurate testing of the large-$\gamma$ sector for future works. As the computational complexity dramatically increases when $K$ increases, it is likely that different numerical methods or approximations would be needed in order to test this sector, which is interesting from a theoretical point of view but however does not appear interesting physically, given the current status of the theory.
\section{Conclusions}
\label{sec:conclusions}
In this work we applied new computational techniques to the study of the infrared divergence of the spinfoam self-energy graph in the EPRL model. We have computed the divergent amplitudes with a running cutoff under an approximation (the ``shells approximation'') which we then lifted by introducing an extrapolation scheme suited for the case considered. We tested our assumptions by refining the approximation up to a large number of shells, using considerable computational resources, for the most interesting case with the Barbero-Immirzi parameter $\gamma = 0.1$. We also explored the divergence at other values of $\gamma$ and investigated the dependence and reliability of our results by using different boundary data and by computing geometrical boundary observables.

Our findings considerably refine the upper and lower bounds given in the previous literature \cite{Riello:2013bzw, Dona:2018infrared}. We have shown that the numerical evidence clearly points to a \emph{linear} divergence in the infrared cutoff parameter.
On one hand, our result shows that the divergence scaling is much less severe than what expected by approximate power-counting arguments of BF and Euclidean theories \cite{Perini:2008pd, Krajewski:2010yq} or upper-bound estimates to the Lorentzian theory \cite{Dona:2018infrared}. On the other hand, our result also shows that the total divergence is of higher order than the analytical estimate of the so-called ``non-degenerate'' sector \cite{Riello:2013bzw} and thus provides the first strong evidence that virtual ``degenerate'' (i.e.\ of non-maximal dimension) configurations effectively dominate the first order quantum corrections to the bare spinfoam propagator. This is a point that was hinted at in previous works (see especially Appendix C of \cite{Riello:2013bzw}) but left unverified.

As in ordinary quantum field theories, understanding quantum corrections to the bare spinfoam propagator is tightly related to the issue of renormalization, which is still an active area of research in the context of spinfoam and group field theories. Our work fills one of the very first step towards this program by providing a precise estimate of the degree of divergence. Our results about the divergence at various values of $\gamma$ also suggest that a possible renormalization scheme might involve the running of the Barbero-Immirzi parameter. The numerical testing of this hypothesis is left for future work, as the possible generalization of our extrapolation scheme to different divergent graphs. 

This is one of the first works in the field of covariant loop quantum gravity to specifically exploit high-performance codes on a computer cluster. The computations accomplished in this paper would not have been possible using a single or a few machines. We believe that complex computational projects will prove to be more and more fundamental to the advancement of the field in the coming years and we hope that our work and our codes will provide a useful and encouraging ground for progressing in this direction. In this regard, the application of MCMC methods in covariant LQG to study the
quantum regime in spinfoams composed of multiple vertex amplitudes glued together will appear soon \cite{Cosm_project}.%
\\[2em]

\paragraph*{\bf Acknowledgments}
We thank Pietro Don\`a for numerous contributions, especially during the early stages of this work. We are grateful to Carlo Rovelli for discussing at length this manuscript and providing many suggestions that improved it.   
The Centre de Calcul Intensif d'Aix-Marseille and the Shared Hierarchical Academic Research Computing Network (SHARCNET) are acknowledged for granting access to their high-performance computing resources. We thank in particular the Compute/Calcul Canada staff for the constant support provided with the Cedar and Graham clusters.
This work was supported by the Natural Science and Engineering Council of Canada (NSERC) through the Discovery Grant "Loop Quantum Gravity: from Computation to Phenomenology". We acknowledge support also from the QISS JFT grant 61466.  FV's research is supported by the Canada Research Chairs Program.
We acknowledge the Anishinaabek, Haudenosaunee, L\=unaap\'eewak and Attawandaron peoples, on whose traditional lands Western University is located.%
\\[2em]
\centerline{***}

\appendix
\section{Consistency checks}
\subsection{Divergence with coherent boundary states}
\label{subsec:div_coherent_state}
In this Section we test whether the results presented in Section \ref{sec:Div_analysis} depend on the chosen boundary data. The basis chosen diagonalizes one of the two dihedral angles of the boundary tetrahedra, thus leaving the other maximally spread. Would the results change using more semiclassical boundary state? To answer this question, we have repeated the computation using intrinsic coherent intertwiners \cite{Rovelli:2014ssa} as boundary states. 

The definition of the intrinsic semiclassical intertwiners is based on the relation between polyhedra and  $SU(2)$ invariants \cite{Barrett:2009gg, article:Dona_etal_2018_SU2_graph_invariants} and coherent intertwiners \cite{Livine:2007vk}. A $4$-valent coherent intertwiner is defined by the group averaging:
\begin{equation}
\label{eq:LivineSpeziale}
||j_a,\vec n_a \rangle \equiv \int dg \bigotimes_{a=1}^4 g\ket{j_a,\vec n_a} = \sum\limits_{i}c_{i}(\vec n_a) |j_a, i \rangle ,
\end{equation}
where $\ket{j_a,\vec n_a}$ are $SU(2)$ coherent states, $g \in SU(2)$ The vectors $\vec n_a$ can be parametrized as $\vec n \equiv (\sin\Theta\cos\Phi, \sin\Theta\sin\Phi,\cos\Theta)$. The decomposition of a $4$-valent coherent intertwiner, in the recoupling channel $i$ with outgoing links, reads
\begin{align}
\label{eq:coeffCS}
c_{i}(\vec n_a) &\equiv \bra{j_a,i}j_a, \vec n_a\ra 
\\ \nn
&= \sum_{m_a} \Wfour{j_1}{j_2}{j_3}{j_4}{m_1}{m_2}{m_3}{m_4}{i} \prod_{a=1}^4
D^{j_a}_{m_a,j_a}(\vec{n}_a)
\\ \nn
&=  \raisebox{-1cm}{\includegraphics[width=4cm]{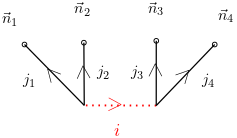}} \ ,
\end{align}
where $D^{j}_{m,j}(\vec{n})=D^{j}_{m,j}(\phi, \Theta,-\phi)$ is the $SU(2)$ Wigner matrix parametrized as:
\begin{equation}
\nn
D^{j}_{m,j}(\phi, \Theta,-\phi)= e^{-i m \phi} d^j_{m,j}(\Theta) e^{i j \phi} \ ,
\end{equation} 
with $d^j_{m,j}$ as a small Wigner matrix \cite{book:varshalovic}. \\
In order to be interpreted as quantum polyhedra, the boundary tetrahedra of the spinfoam must satisfy the closure condition:
\begin{equation}
\label{eq:closure_condition}
  \sum_f \Lambda_{f} \hspace{1mm} \vec{n}_{f}  = 0
\end{equation}
where $\Lambda_{f}$ is the eigenvalue of the area operator associated with each face of the tetrahedron. 

\begin{figure*}[!btp]
   \centering
   \includegraphics[width=0.3\linewidth]{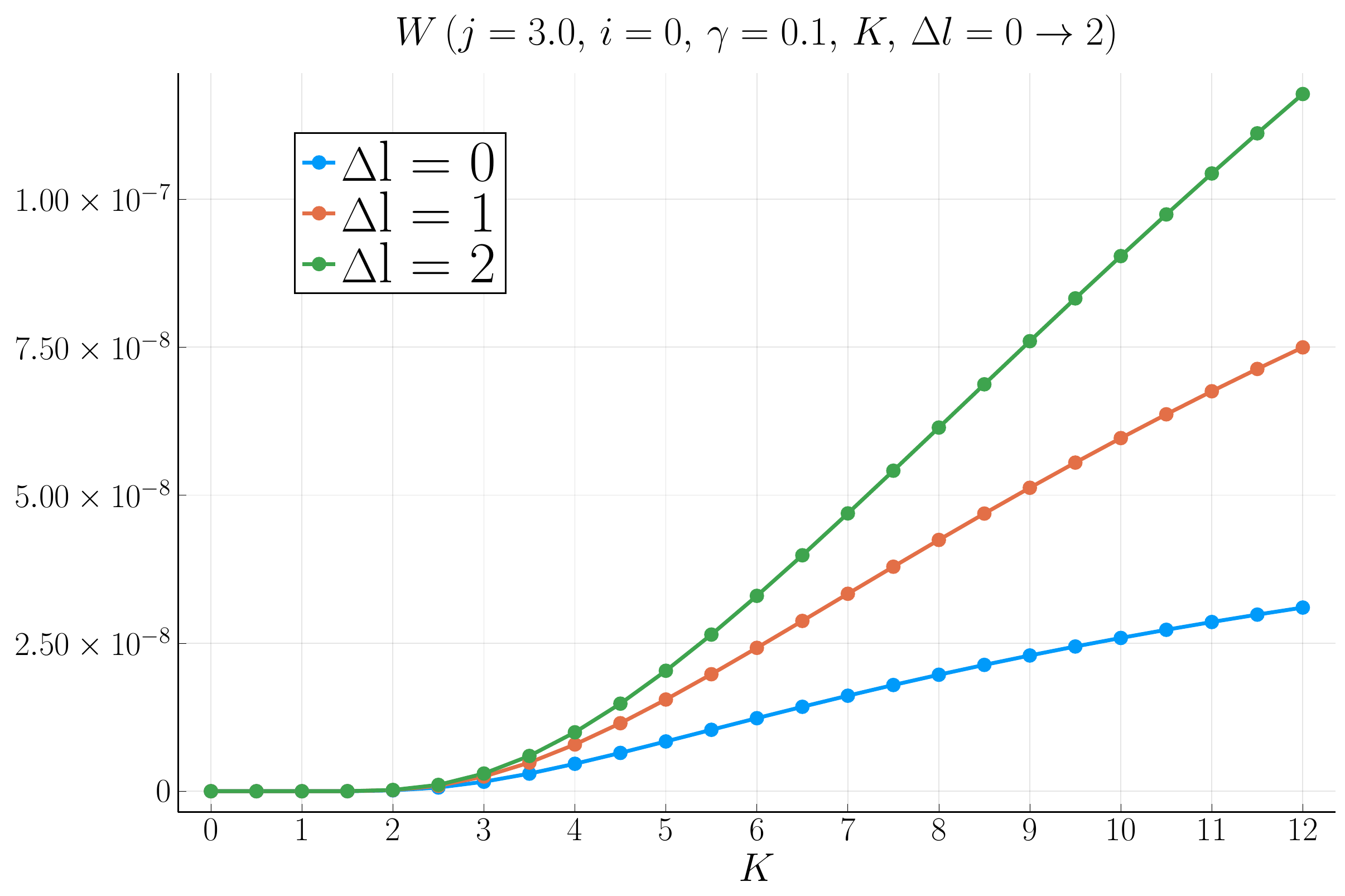}
   \includegraphics[width=0.3\linewidth]{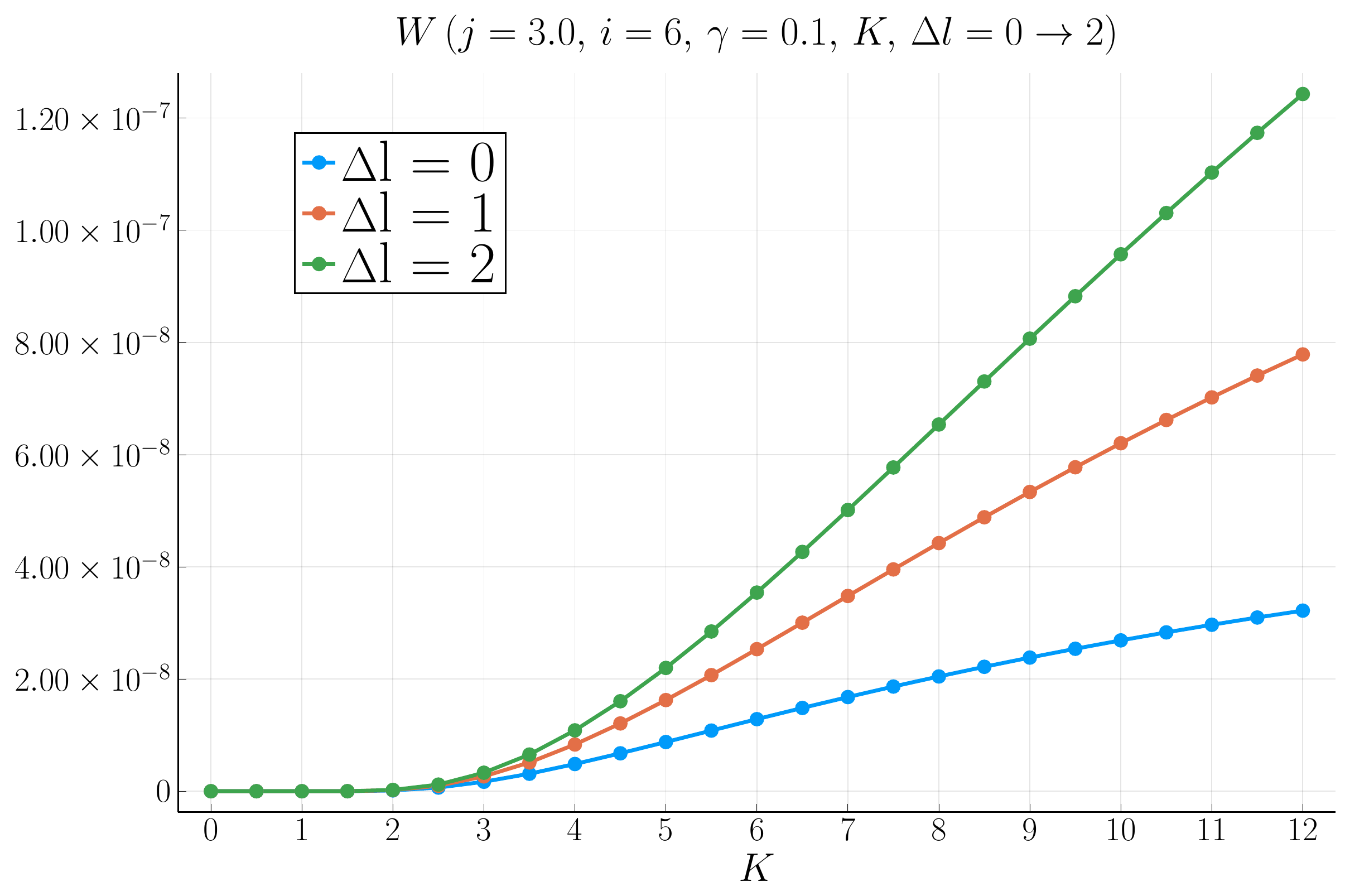}
      \includegraphics[width=0.3\linewidth]{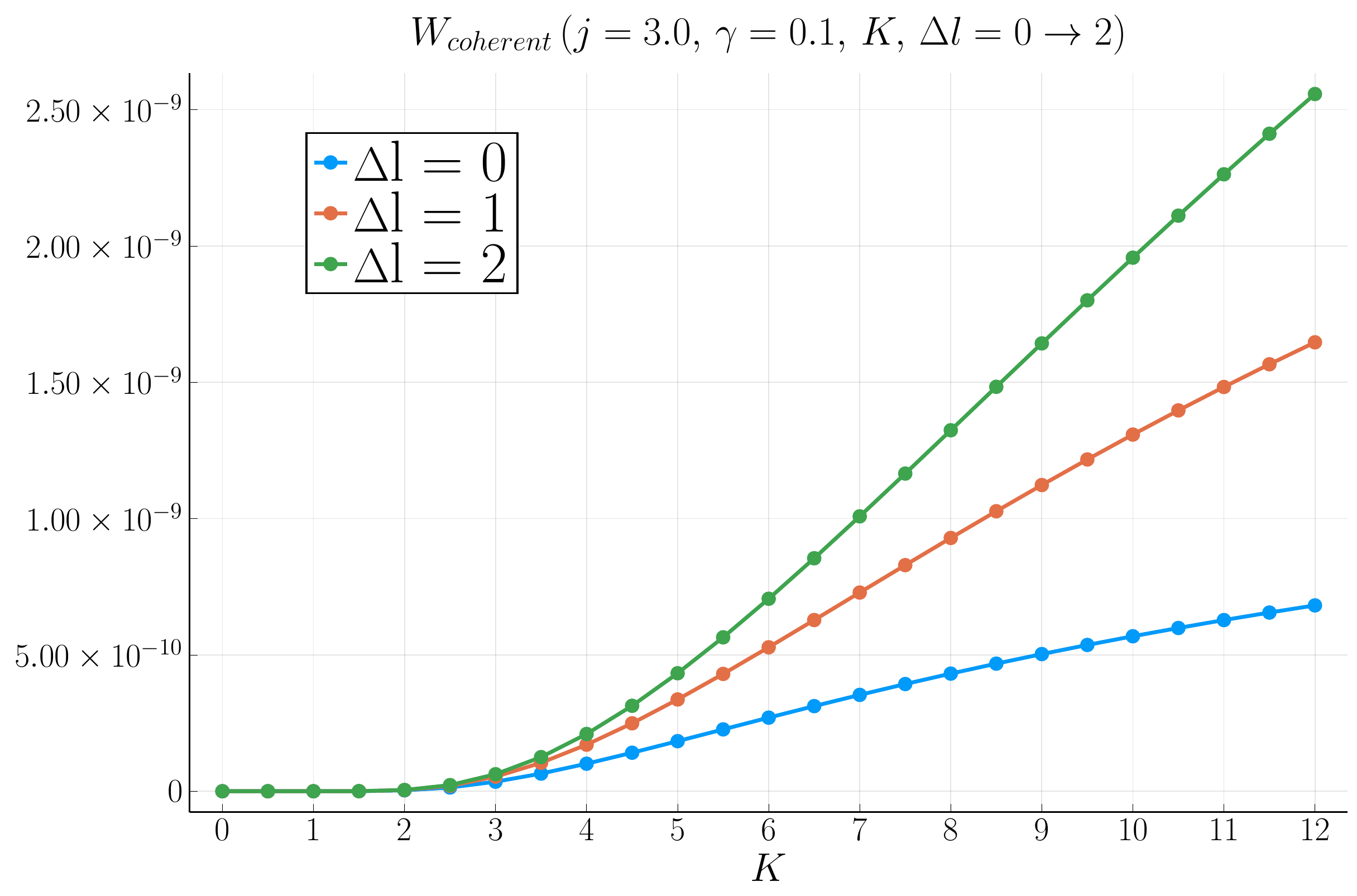}
    \caption{ 
    \label{fig:increasing_boundary_scaling} 
    Self-energy amplitude \eqref{eq:EPRL_self_energy_amplitude} with fixed intertwiners (first two panels) versus coherent (third panel) boundary data \eqref{eq:LivineSpeziale}. Using coherent intertwiners, we find no difference in scaling by changing the orientation of the normals $\vec{n}_{a}$.
    }
\end{figure*}

An example of the numerical analysis of the amplitude obtained by contracting the amplitude \eqref{eq:EPRL_self_energy_amplitude} with coherent states \eqref{eq:LivineSpeziale}, for increasing boundary spin values, is shown\footnote{normals $\vec{n}_a$ satisfy condition \eqref{eq:closure_condition} and we considered the absolute value in order to obtain real numbers} in Figure \ref{fig:increasing_boundary_scaling}, compared with boundary spinnetwork states. The scaling of the self-energy amplitude turns out to be unchanged whether using boundary spinnetwork states $|j,i_{\pm} \rangle$ or coherent states $||j,\vec n_{\pm f} \rangle$, apart for a global factor that multiplies all curves. Also, using coherent states, the scaling does not seem to depend on the orientation of the normals $\vec n_{\pm f} $.

\subsection{Boundary observables} 
\label{sec:Boundary_observables}
Another check of consistency of the geometrical picture and the numerical calculation can be obtained by looking at the geometry of the two boundary tetrahedra. Their geometrical properties are jointly fixed by choosing the boundary data and by the dynamics. 
Here  
we show  
the resulting 
 expectation  
  values of  
  some geometrical boundary operators, that is: 
\begin{equation}
\label{eq:dynamic_exp_value}
\langle O \rangle =  
\frac
{ \langle W | O | \Psi \rangle }  
{ \langle W | \Psi \rangle }  
\end{equation}
where the bra $W$ contains the propagator, namely the dynamics, while the ket $\Psi$ turns out to be the tensor  
product  
of the  
\textit{ in } and \textit{ out } states  
of the LQG  
Hilbert  
space. With the term "propagator" we 
 refer to the square matrices  
(they are  
such since the  
self-energy triangulation  
 has two boundary 
 tetrahedra) in  
which the  
element $a, b$ corresponds to the amplitude \eqref{eq:EPRL_self_energy_amplitude} with $a = i_+, b = i_{-}$. Computing the expectation values \eqref{eq:dynamic_exp_value} of geometric operators describing boundary tetrahedra is a significant numerical investigation, especially considering that computations  
carried  
out with  
the EPRL  
model are  
still in their  
primordial  
stages \cite{Dona2019, article:Dona_etal_2019_numerical_study_lorentzian_EPRL_spinfoam_amplitude, Gozzini_primordial, Spinfoam_on_a_Lefschetz_thimble, han2021complex, frisoni2021studying}, and, as far as we know, at present time there are no such numerical computations with spinfoams containing a bubble. 

Since in the calculation of \eqref{eq:dynamic_exp_value} the dimensional factor of the boundary intertwiners is relevant, in this section it is convenient to define a "normalized" amplitude as:
\begin{equation}
\label{eq:normalized_ampl}
W_N(j, i_{\pm}, K) \equiv W (j, i_{\pm}, K) \sqrt{d_{\{i_{\pm} \}}} \ ,
\end{equation}
where $d_{\{i_{\pm}\}}=\prod\limits_{\pm}(2 i_{\pm} +1) = (2 i_{+} +1) (2 i_{-} +1)$.
\subsubsection{Angles}
\label{subsec:angles}
\begin{figure}[!hbtp]
    \centering
        \includegraphics[width=8.7cm]{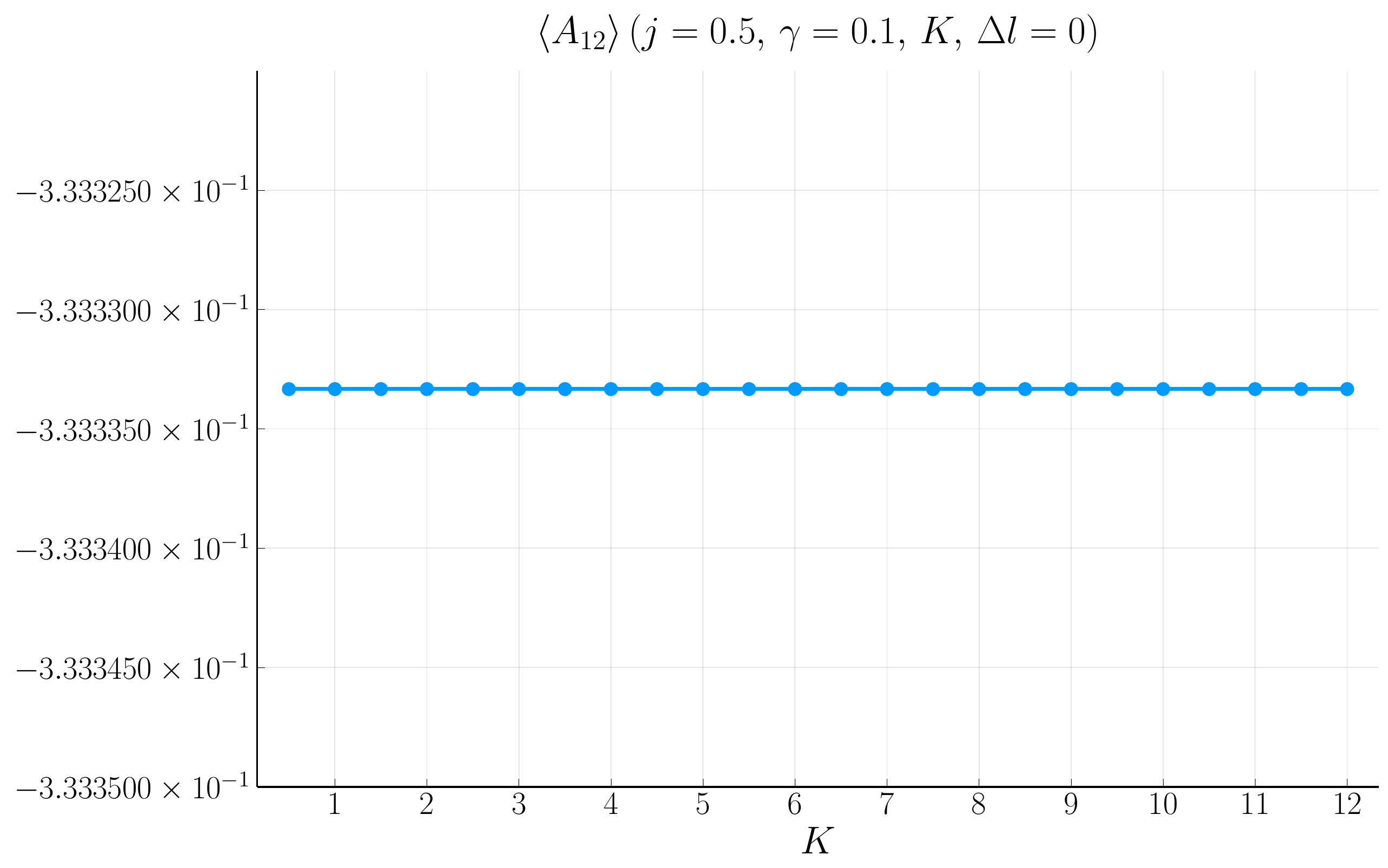}
    \caption{\label{fig:angles} 
    Dynamic expectation value of the cosine of the external dihedral angle operator \eqref{eq:dynamic_angle}. We find an excellent agreement with the value $-0.\Bar{3}$. }
\end{figure}

The shape 
 of the  
 tetrahedra  
 in twisted  
 geometry  
 is measured  
 by the angle  
operator:
\begin{equation}  \label{eq:geom-angleop}
  A_{ab} |j, i_{\pm} \rangle 
  =
   \cos( \theta_{ab} )_{i_{\pm}} 
   |j, i_{\pm} 
   \rangle
\end{equation}
which is interpreted as the cosine of the external dihedral angle between the faces $a$ and $b$ of the tetrahedron defined on the node associated with the intertwiner $i_{\pm}$. The spin-network basis states diagonalize the dihedral angle $\theta_{ab}$ between faces $a$ and $b$. The equation for the dihedral angle $\cos(\theta_{ab})$ in terms of intertwiner spin $i_{\pm}$ reads:
\begin{equation}.  \label{eq:geom-angleformula}
  \cos( \theta_{ab} )_{i_{\pm}} 
  = 
  \frac
  {  i_{\pm}   (i_{\pm} +1  ) - j_a(j_a+1) - j_b(j_b+1)}
  {2\sqrt{j_a(j_a+1)j_b(j_b+1)} }.
\end{equation}
We consider 
 the expectation 
 value \eqref{eq:dynamic_exp_value}  
of the  
angle operator  
\eqref{eq:geom-angleop}  
in any  
of the two  
(equal) boundary  
regular tetrahedra 
 of the triangulation  
using the spinnetwork  
state. According 
 to the  
recoupling  
basis  
$ ( j_1 , j_2 ) $,  
we focus  
on the  
angle  
between  
faces $ 1 $  
and $ 2 $. The  
expectation  
value on node $+$ can be computed as: 
\begin{equation}
\frac{\langle W | A_{12} | W \rangle}{ \langle W  | W \rangle } = \frac{ \sum\limits_{i_{\pm}} \left[ W_N(j, i_{\pm}, K) \right]^2 \cos(\theta_{12})_{i_+}}{ \sum\limits_{i_{\pm}} \left[ W_N(j, i_{\pm}, K)  \right]^2},
\label{eq:dynamic_angle}
\end{equation}
where we considered the case in which all boundary spins are equal to $j$. 
Carrying  
out the  
numerical computation  
of \eqref{eq:dynamic_angle},  
we obtain  
a value  
that is  
in agreement  
with the  
geometric  
value  
of the  
external  
angle  
of a  
regular  
tetrahedron  
up to the  
tenth significant  
digit, as  
shown  in   Figure 
 \ref{fig:angles}. We repeated the calculation for increasing values of boundary spins $j$ up to $j = 5$, also varying $\Delta l$, finding exactly the same value and the unaltered precision.
This is  
consistent  
with the  
fact that  
we are  
looking at  
the only  
angle which  
is completely  
sharp,  
while the  
others  
turn out  
to be  
spread. 
\subsubsection{Volumes}
\label{subsec:volumes}
\begin{figure}[!b]
    \centering
        \includegraphics[width=.85\linewidth]{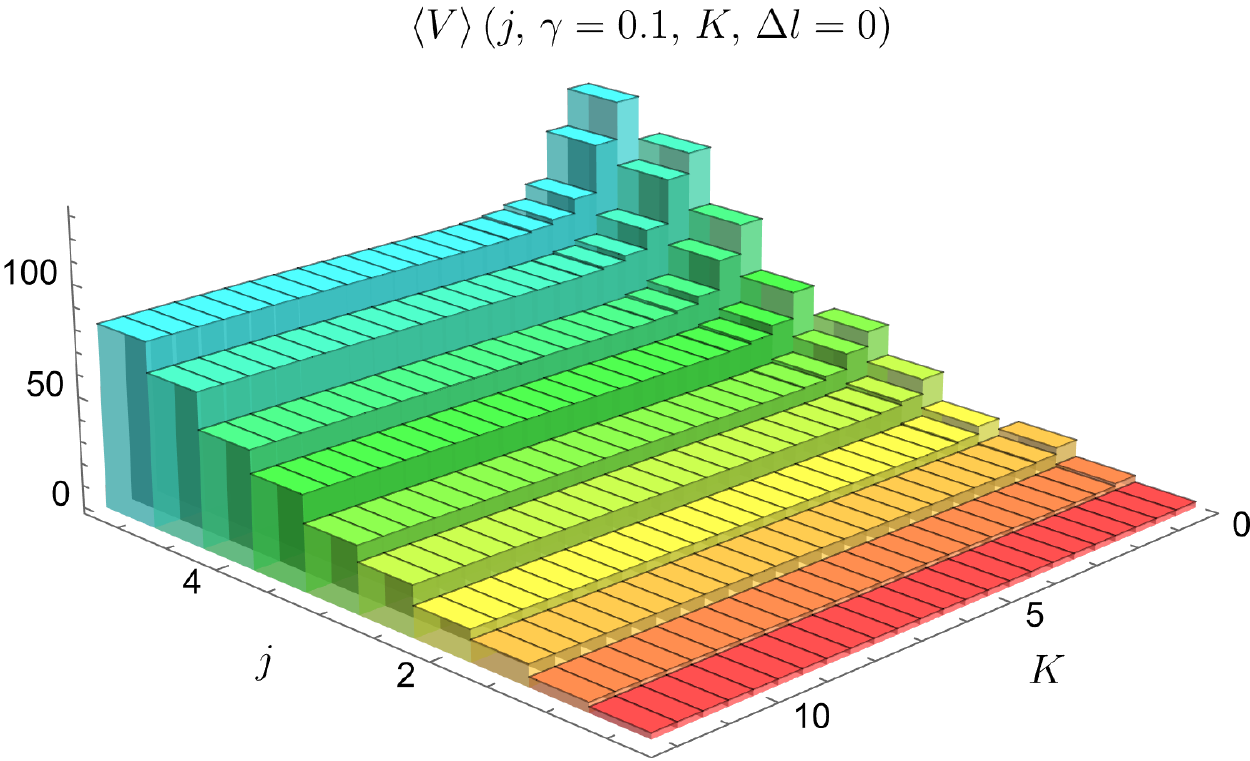}
    \caption{\label{fig:volumes} 
    Dynamic expectation value of the volume operator \eqref{eq:dynamic_volume}. We neglected all the constant factors in \eqref{eq:dynamic_volume} and \eqref{eq:volume_eigenvalues}. The volumes scale as $j^{\frac{3}{2}}$.}
    \end{figure}

The volume operator of a tetrahedron reads:
\begin{equation}
\label{eq:volume_operator}
V_{abc} = \frac{\sqrt2}{3}\left(8 \pi G \hbar \gamma \right)^{\frac{3}{2}}\sqrt{\abs{\vec{J}_a \cdot \vec{J}_b \times \vec{J}_c}} \ ,
\end{equation}
where $\vec{J}_a$ is the vector of $SU(2)$ generators on link $a$.
The operator \eqref{eq:volume_operator} is not diagonal in the spinnetwork states. The dynamic expectation value (see \cite{Rovelli:2014ssa} for the derivation) on node $+$ turns out to be:
\begin{align}
\label{eq:dynamic_volume}
& \frac{\langle W | V | W \rangle}{ \langle W  | W \rangle } = \\
&\nn  =  \frac{ \sum\limits_{i_{\pm},i'_{+}} \sqrt{ d_{\{ i_{+} \}} d_{ \{ i_{+}' \}} }   W_N(j, i_{\pm}, K) W_N(j, i_{+}', K)  v \left( i_{+} , i_{+}', j \right)}{  \sum\limits_{i_{\pm}} \left[ W_N(j, i_{\pm}, K) \right]^2 }
 \end{align}
where $v \left( i_+ , i_+', j \right)$ is the matrix element:
\begin{equation}
\label{eq:volume_eigenvalues}
v \left( i_+ , i_+', j \right) = \frac{\sqrt{2}}{3} \left( 8 \pi G \hbar \gamma \right)^{\frac{3}{2}} \sum\limits_v \langle j, i_+' | v \rangle \sqrt{ |q_v| }  \langle  v | j, i_+ \rangle,  
\end{equation}
in which $ q_v $ are the eigenvalues of the volume, while $ | v \rangle $ corresponds to the eigenvector relative to the eigenvalue $ q_v $. In Figure \ref{fig:volumes} we report the numerical values obtained from \eqref{eq:dynamic_volume}. We see that the scaling of the expectation value of the boundary volume as a function of the spin $j$ corresponds to that actually existing between the volume of a regular tetrahedron and the area of one of its faces, that is $V \propto j^{3/2}$. As for the angle operator, we repeated the calculation by varying the number of shells, finding the same trend.

Finally, both for the angle operator and for the volume, we repeated the calculation with boundary coherent states \eqref{eq:LivineSpeziale} verifying that the expectation value is dominated by the classical contribution.

The picture that emerges is that the boundary observables have an expectation value that coincides with the classical one, and are not affected by the presence of divergence in the bulk of the spinfoam. Namely, even if the amplitude \eqref{eq:EPRL_self_energy_amplitude} is divergent, the expectation values \eqref{eq:dynamic_angle}, \eqref{eq:dynamic_volume} are finite and fully consistent with the geometry of a regular tetrahedron.
\begin{figure*}[!t]
  \begin{subfigure}[b]{0.3\textwidth}
        \includegraphics[width=\linewidth]{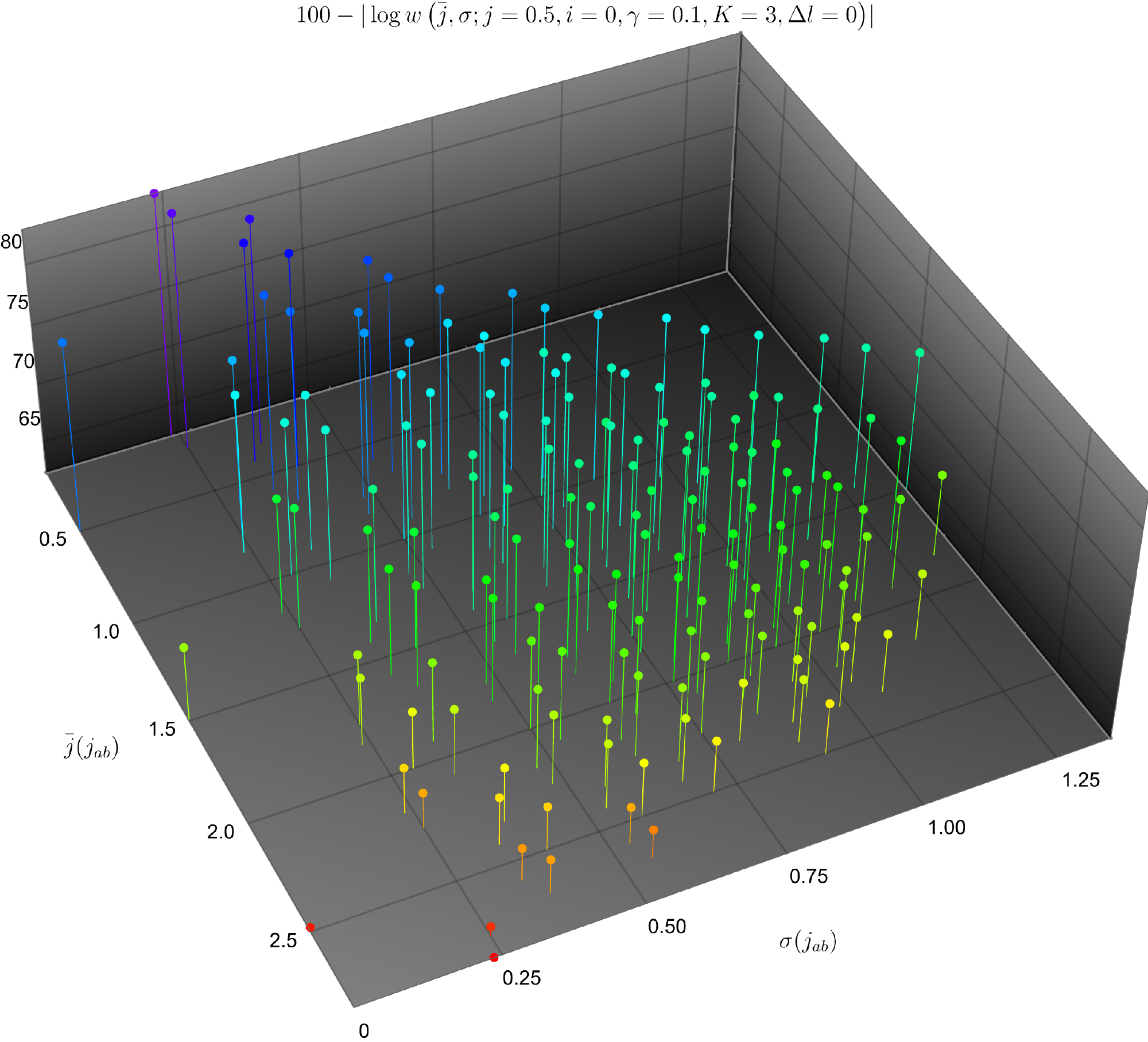}
    \end{subfigure}
    \begin{subfigure}[b]{0.3\textwidth}
        \includegraphics[width=\linewidth]{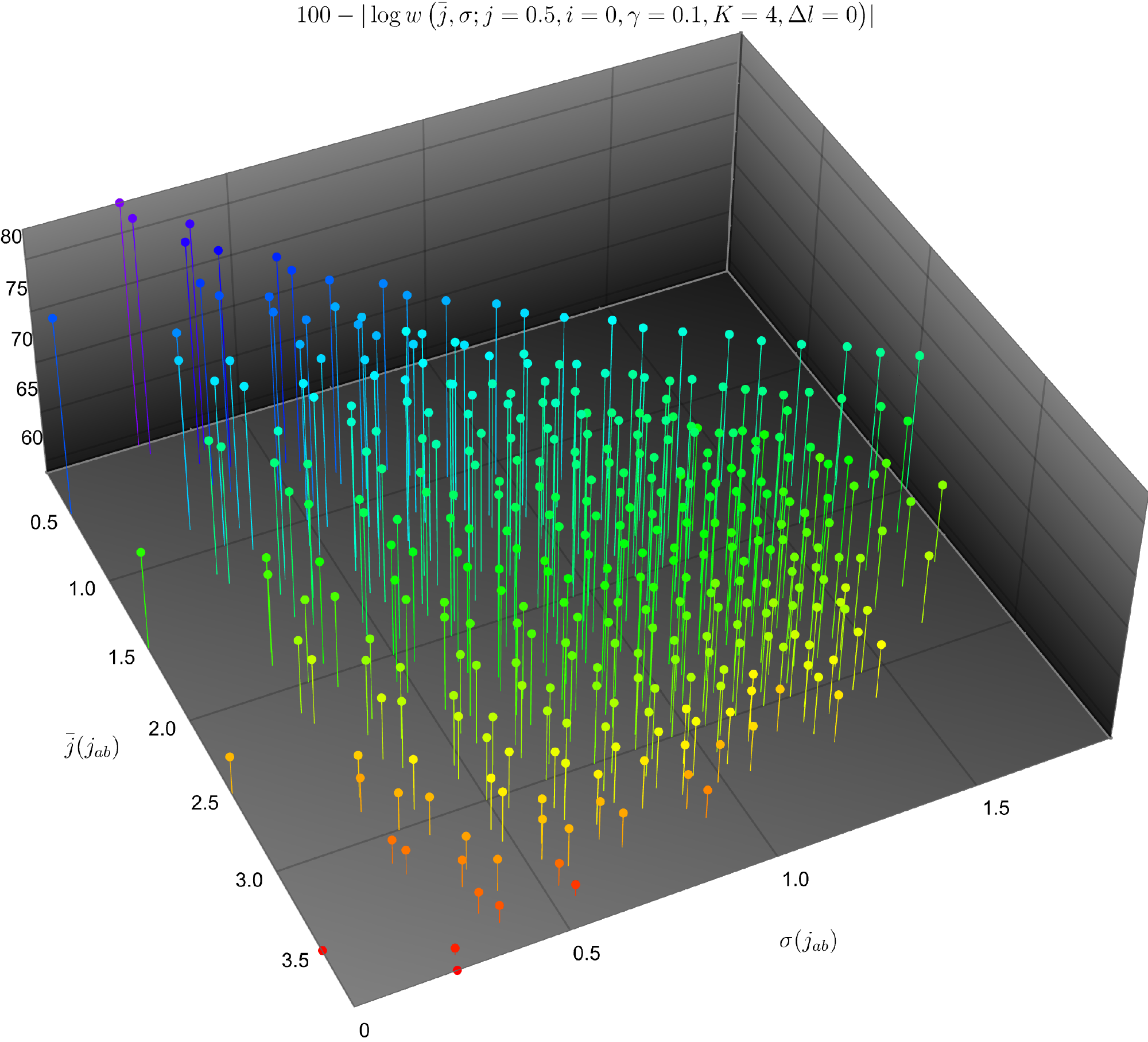}
    \end{subfigure}
  \begin{subfigure}[b]{0.3\textwidth}
        \includegraphics[width=\linewidth]{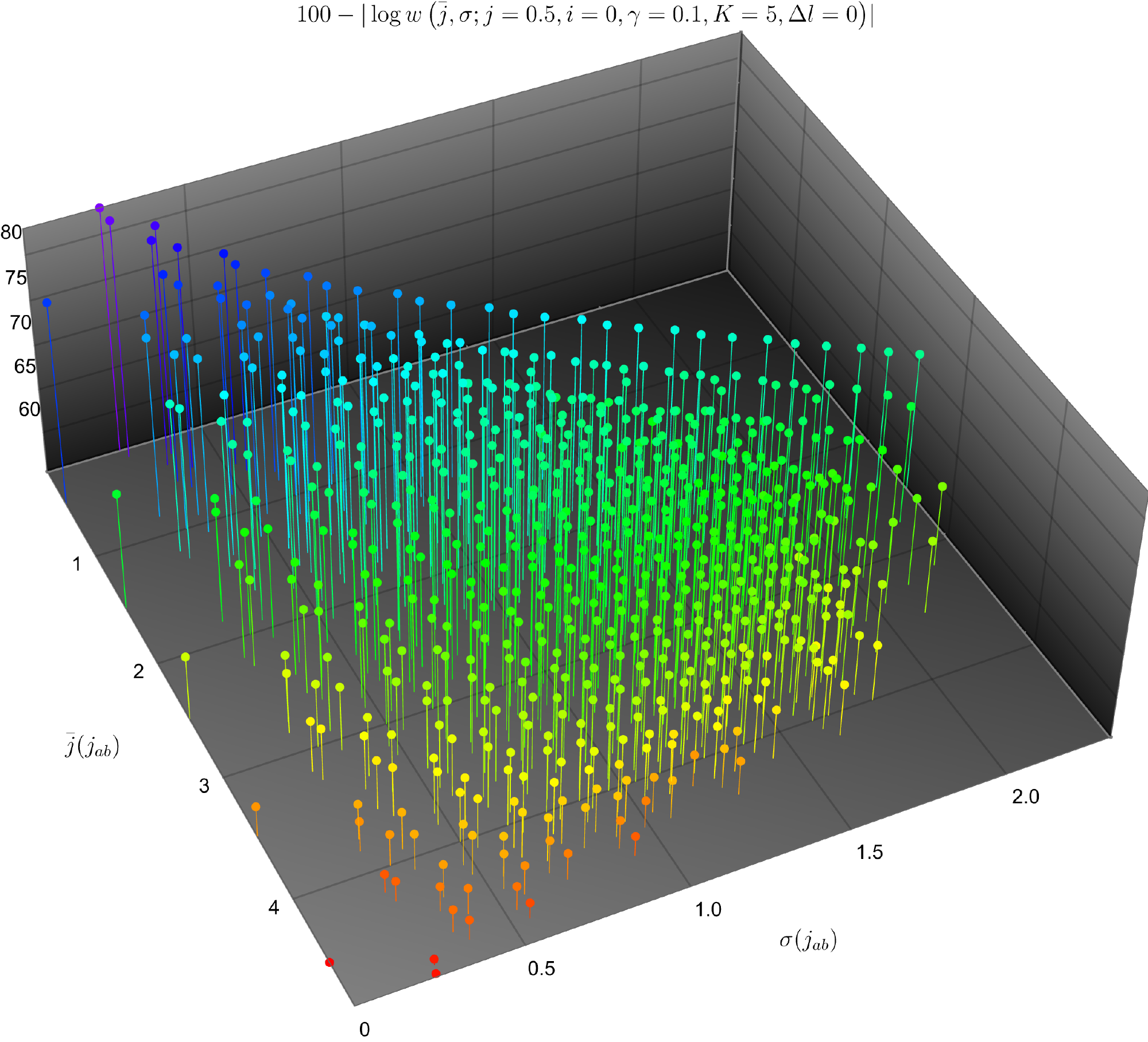}
    \end{subfigure}    
    \caption{ \label{fig:internalcontributes} 
    The largest individual contributions to the divergent amplitude \eqref{eq:EPRL_self_energy_amplitude_compact} come from ``semiclassical'' bulk spins configurations with $\Bar{j}\sim j$ and low $\sigma$. Triangular inequalities result in a triangular shape in the plane $\left( \Bar{j}, \sigma \right)$.} 
\end{figure*}

\section{Internal contributions}
\label{app:internal_contributions}

We add here a tentative analysis of the kind of configurations that contribute to the divergence. For this aim, it is convenient to rewrite equation \eqref{eq:EPRL_self_energy_amplitude_compact} as
\begin{equation}
\label{eq:EPRL_internal_contributions}
W_{} \left( j, i, \gamma; K , \Delta l \right) = \sum\limits_{j_{ab} \leq K} \omega(j_{ab}; j, i, \gamma, \Delta l).  
\end{equation} 
Now, consider the average and the standard deviation of the sextuplet of internal spins $\{j_{ab}\}$:
\be
\Bar{j}(j_{ab})=\frac16 \sum_{a\ne b}j_{ab},\ \  \  \sigma(j_{ab})= \sqrt{\frac16 \sum_{a\ne b} (j_{ab}-\Bar{j}_{ab})^2}
\ee
and define 
\be
w(\Bar{j},\sigma ; j, i, \gamma , K, \Delta l) = \sum\limits_{\substack{\Bar{j}(j_{ab})=\Bar{j}\\ 
\sigma(j_{ab})=\sigma}} \omega(j_{ab}; j, i, \gamma, \Delta l).  
\ee

\noindent That is, we write the contributions to the divergence as a function of the average and standard deviation of the internal spins, adding the contributions with different sextuplets that have the same average and standard deviation. The function $100 - \abs{\log{w(\Bar{j},\sigma; j, i, \gamma , K, \Delta l}})$ is plotted for fixed  $j, i, \gamma, \Delta l$ and various values of $K$ in Figure \ref{fig:internalcontributes}, where the factor $100$ is only for visual purposes.
Increasing $\Delta l$ has the effect of increasing the value of each point, while increasing K also implies adding more configurations, as shown in Figure \ref{fig:jab-number}. Figure \ref{fig:internalcontributes} indicates that the largest individual contributions to the amplitude \eqref{eq:EPRL_self_energy_amplitude_compact} come from ``semiclassical'' bulk spins configurations with $\Bar{j}\sim j$ and low $\sigma$, as one might expect. While the divergence comes from the proliferation of smaller contribution with arbitrary high spin and deviations.

We also note that the divergence of the self-energy, manifestly shown in Figure \ref{fig:increasingscaling}, occurs in the limit $\Delta l \rightarrow \infty$ only. In fact, in \cite{Dona:2018infrared} it is shown that the sum over the internal spins of the amplitude \eqref{eq:EPRL_self_energy_amplitude_compact} is convergent in the simplified model $\Delta l = 0$. Even if an analytical proof is not yet available, it seems likely that the amplitude has a finite limit $K \rightarrow \infty$ for each finite $\Delta l$, since the internal spins configurations in the limit become more and more suppressed. Since also the limit $\Delta l \rightarrow \infty$ at fixed $K$ is finite, we can schematize the role of the two limits with the following four node diagram:
\[
\begin{tikzcd}
W(K, \Delta l)  \arrow{r}{K \rightarrow \infty} \arrow[swap]{d}{\Delta l \rightarrow \infty} & \text{convergence?} \arrow{d}{\Delta l \rightarrow \infty} \\
\text{convergence} \arrow{r}{K \rightarrow \infty} & \text{divergence} .
\end{tikzcd}
\]
\\

\section{Conventions and basic formulas} 
\label{app:conventions}
We collect here the basic definitions needed for the calculation in this paper. 
\subsection{SU(2) Wigner j-symbols} 
\label{app:SU(2)_symbols}
We use the definition of the Wigner's $(3jm)$ symbol given in \cite{book:varshalovic} with the orthogonality property
\begin{equation}
\sum_{m_1,m_2}\!\! \Wthree{ j_1}{ j_2}{ j_3} {m_1}{m_2}{m_3}\!\! \Wthree{ j_1}{ j_2}{ j_3} {m_1}{m_2}{n_3} = \frac{\delta_{j_{3}l_{3}}\delta_{m_3n_3}}{2 j_3 + 1}. 
\end{equation}
These are normalized and vanish if triangular inequalities are not satisfied. We define $(4jm)$ symbol as the contraction of two $(3jm)$ symbol using an intertwiner $k$:
\begin{align}
\label{eq:4jm_symbol}
&i^{(k)}_{m_1 m_2 m_3 m_4} \equiv \Wfour{ j_1}{ j_2}{ j_3} {j_4}{m_1}{m_2}{m_3}{m_4}{k}\\[1em] \nn
& \equiv \sum_{m_i} (-1)^{k-m_i} \Wthree{ j_1}{ j_2}{ k} {m_1}{m_2}{m_i} \Wthree{ k}{ j_3}{ j_4} {-m_i}{m_3}{m_4}\ ,
\end{align}
with the orthogonality relations:
\begin{align}
&\sum_{m_1,m_2,m_3} \Wfour{ j_1}{ j_2}{ j_3} {j_4}{m_1}{m_2}{m_3}{m_4}{k_1} \Wfour{ j_1}{ j_2}{ j_3} {j_4}{m_1}{m_2}{m_3}{m_4}{k_2}  \nn \\[1em] &= \frac{\delta_{k_1 k_2}}{2 k_1 + 1} \frac{\delta_{j_{4}l_{4}}\delta_{m_4n_4}}{2 j_4 + 1} \ ,
\end{align}
where $ \frac{\delta_{k_1 k_2}}{2 k_1 + 1}$ is the normalization factor.
A synthetic notation for $(4jm)$ symbols is also:
\begin{equation}
\left(\begin{array}{c} j_f \\ m_f \end{array}\right)^{(k)} \equiv \Wfour{ j_1}{ j_2}{ j_3} {j_4}{m_1}{m_2}{m_3}{m_4}{k}. 
\end{equation}
With these definitions and the conventional definitions of Wigner's matrices for SU(2), 
$D^{(j)}_{mn}(g)$, we have:
\begin{align}
\label{intD3App}
& \int dg\ D^{j_1}_{m_1n_1}  \left( g \right) D^{j_2}_{m_2 n_2} \left( g \right) D^{j_3}_{m_3 n_3} \left( g \right) D^{j_4}_{m_4 n_4} \left( g \right) \\[1em] \nn &= 
\sum_{k} d_{k} \Wfour{j_1}{j_2}{j_3}{j_4}{m_1}{m_2}{m_3}{m_4}{k} \Wfour{j_1}{j_2}{j_3}{j_4}{n_1}{n_2}{n_3}{n_4}{k}.
\end{align}
We denote the 4-valent intertwiner with the ket $| k \rangle$, defined as:
\begin{equation}
\label{eq:intertwiner_non_normalized_definition}
| k \rangle \equiv  \hspace{1mm} i^{(k)}_{m_1 m_2 m_3 m_4} | j_1,m_1 \rangle | j_2,m_2 \rangle  | j_3,m_3 \rangle  | j_4,m_4 \rangle 
\end{equation}
where the $(4jm)$ symbol is defined in \eqref{eq:4jm_symbol}. The states \eqref{eq:intertwiner_non_normalized_definition} constitute a basis in the intertwiner space of a 4-valent node:
\begin{equation}
\label{eq:intertwiner_space}
\mathcal{I}_4 \equiv Inv \left[ V^{j_1} \otimes V^{j_2} \otimes V^{j_3} \otimes V^{j_4} \right]   
\end{equation}
where $V^{j_i}$ is the irreducible representation of spin $j_i$. The resolution of the identity in the intertwiner space $\mathcal{I}_4$ and the orthogonality condition can be written as:
\begin{align}
\label{box:relations_non_normalized_intertwiner}
\mathbf{1}^{( \mathcal{I}_4 ) } & = \sum_{k = 0}^{2j} (2k + 1) | k \rangle \langle k |, \\ 
\langle k | k' \rangle & = \frac{\delta_{k,k'}} {(2k + 1)} .
\end{align}
The 15-j symbol we use in this work is the irreducible symbol of first type, following the convention of \cite{GraphMethods}. Its graphical representation and its definition in terms of Wigner's 6-j symbols is%
\begin{widetext}
\begin{align}
\label{eq:15jconv}
&
\hspace{2cm}
\raisebox{-15mm}{ \includegraphics[width=4.0cm]{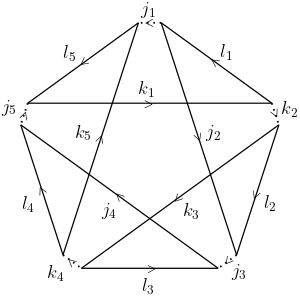}} ~=~ \left \{ \begin{array}{ccccc} j_1 & j_2 & j_3 & j_4 & j_5 \nn \\
l_1 & l_2 & l_3 & l_4 & l_5 \nn \\ 
k_1 & k_2 & k_3 & k_4 & k_5 \end{array}\right \}  ~=&  \\[1.5em]
&
=~  (-1)^{\sum_{i=1}^5 j_i + l_i +k_i} \sum_s d_s \Wsix{j_1}{k_1}{s}{k_2}{j_2}{l_1} \Wsix{j_2}{k_2}{s}{k_3}{j_3}{l_2} 
\Wsix{j_3}{k_3}{s}{k_4}{j_4}{l_3} \Wsix{j_4}{k_4}{s}{k_5}{j_5}{l_4} \Wsix{j_5}{k_5}{s}{j_1}{k_1}{l_5}  \ .
&
\end{align}
\end{widetext}

\subsection{Booster functions} 
\label{app:booster}
The booster functions \cite{Dona2018}, \cite{article:Dona_etal_2019_numerical_study_lorentzian_EPRL_spinfoam_amplitude}, also known as B4 functions \cite{Speziale2016}, are the non compact remnants of the $SL(2, \mathbb{C})$ integrals. These functions turn out to encode all the details of the EPRL model, such as the $Y -$map. We define them as:
\begin{widetext}
\begin{equation}
\label{eq:boosterdef}
B_4 \left( j_f, l_f ;  i ,  k \right) \equiv \frac{1}{4\pi} \sum_{ p_f } \left(\begin{array}{c} j_f \\ p_f \end{array}\right)^{(i)} \left(\int_0^\infty \mathrm{d} r \sinh^2r \, 
\prod_{f=1}^4 d^{(\gamma j_f,j_f)}_{j_f l_f p_f}(r) \right)
\left(\begin{array}{c} l_f \\ p_f \end{array}\right)^{(k)}\  = \raisebox{-12.5mm}{ \includegraphics[width=2.4cm]{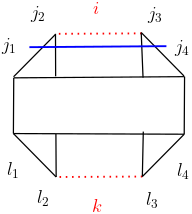}},
\end{equation}
\end{widetext}
where $d^{(\rho,k)}(r)$ are the boost matrix elements for $\gamma$-simple irreducible representations of $SL(2,\mathbb{C})$ in the principal series, $\gamma$ is the Barbero-Immirzi parameter and the $(4jm)$ symbols are defined in section \ref{app:SU(2)_symbols}.

The booster  
function  
$ B_4 \left(  j_f, l_f ;  i ,  k  \right) $ 
is interpreted  
as a  
quantum   
tetrahedron  
being  
boosted  
among  
adjacent  
frames:  
the two  
sets $j_f$ and 
 $l_f$ describe 
 the four areas 
 of the  
tetrahedron 
 in the  
two frames  
connected  
by a boost,  
and the  
two intertwiners  
$i$ and $k$  
describe the  
quantum 
intrinsic  
shape  
of the  
tetrahedron \cite{Speziale2016}. For a precise interpretation of the booster functions and their semiclassical limit we refer to \cite{dona2020}.
The explicit form of the boost matrix elements can be found in the literature. For the general form, see \cite{Ruhl:1970fk, Speziale2016}. We just report here the case of simple irreducible representations:
\begin{widetext}
\begin{align}
\label{eq:dsmall}
d^{(\gamma j,j)}_{jlp}(r) =&  
(-1)^{\frac{j-l}{2}} \frac{\Gamma\left( j + i \gamma j +1\right)}{\left|\Gamma\left(  j + i \gamma j +1\right)\right|} \frac{\Gamma\left( l - i \gamma j +1\right)}{\left|\Gamma\left(  l - i \gamma j +1\right)\right|} \frac{\sqrt{2j+1}\sqrt{2l+1}}{(j+l+1)!}  
\left[(2j)!(l+j)!(l-j)!\frac{(l+p)!(l-p)!}{(j+p)!(j-p)!}\right]^{1/2} \nn \\
&\ \times e^{-(j-i\gamma j +p+1)r}
\sum_{s} \frac{(-1)^{s} \, e^{- 2 s r} }{s!(l-j-s)!} \, {}_2F_1[l+1-i\gamma j,j+p+1+s,j+l+2,1-e^{-2r}] \ .
\end{align}
\end{widetext}
$^{}$\\[4em]


\providecommand{\href}[2]{#2}\begingroup\raggedright\endgroup

\end{document}

%% file: config.tex
\pdfoutput=1

\usepackage[autostyle, english = british]{csquotes}
\MakeOuterQuote{"}

\usepackage[english]{babel}

\usepackage[format=plain,
            textfont=it]{caption}

\usepackage{graphicx}
\usepackage{adjustbox}
\usepackage{subcaption} 
\usepackage{capt-of}
\usepackage{amsmath,amssymb,amsfonts,tensor}
\usepackage{xcolor}
\usepackage{graphicx}
\usepackage{subcaption}
\usepackage{float}
\usepackage{dsfont}
\usepackage{enumitem}
\usepackage[section]{placeins}
\usepackage{empheq}
\usepackage{tikz-cd}

\usepackage{algorithm}
\usepackage[noend]{algpseudocode}

\usepackage{physics} 
\usepackage[percent]{overpic} 

\usepackage[colorlinks,citecolor=blue,linkcolor=blue,urlcolor=blue,pdfencoding=auto, psdextra]{hyperref}

\newcommand{\be}{\begin{equation}}
\newcommand{\ee}{\end{equation}}

